\documentclass{iopart}

\usepackage[font=small,skip=0pt]{caption}
\usepackage[utf8]{inputenc}
\usepackage{pdflscape}
\usepackage{rotating}
\usepackage{tabularx}
\usepackage{rotating,multirow}
\usepackage{longtable}
\usepackage{array,graphicx}
\usepackage{subcaption}
\usepackage{booktabs}
\usepackage{hyperref}
\hypersetup{colorlinks=true,urlcolor=blue,citecolor=blue,linkcolor=blue} 
\newcolumntype{x}[1]{>{\centering\arraybackslash}p{#1}}
\usepackage{xcolor}
\usepackage{graphicx}
\usepackage{dcolumn}

\usepackage[left]{lineno}

\renewcommand{\figurename}{Figure}
\renewcommand{\tablename}{Table}

\begin{document}

\title[Preparing students for the quantum information revolution]{Preparing students for the quantum information revolution: Interdisciplinary teaching, curriculum development, and advising in quantum information science and engineering
}

\author{Fargol Seifollahi$^1$\footnote{Corresponding author},  and Chandralekha Singh$^1$}

\address{$^1$ Department of Physics and Astronomy, University of Pittsburgh, Pittsburgh, Pennsylvania, 15260 USA}

\ead{fas102@pitt.edu}

\date{\today}

\begin{abstract}

As the field of quantum information science and engineering (QISE) continues its rapid growth, there are increasing concerns about the workforce demands and the necessity of preparing students for quantum-related careers. Given the interdisciplinary nature of the field, it is necessary to offer diverse educational opportunities to ensure that students are well-prepared for careers emerging from the second quantum revolution. In this paper, we present our findings from a qualitative study involving semi-structured interviews with university QISE educators, who have taken on the challenges and opportunities in developing QISE courses and curricula for undergraduate and graduate students from different academic backgrounds. Our findings focus on common themes across undergraduate and graduate QISE education, as well as advising and mentoring students to prepare them for research in the field. The interviewees discussed the various strategies they had implemented, such as incorporating hands-on lab activities, integration of Python coding with Qiskit, and including project-based learning experiences. Furthermore, their reflections on mentorship and advising students emphasized the importance of recognizing students' prior preparation, providing targeted resources, and supportive learning environments. Our findings are meant to provide guidance for educators looking to implement effective QISE educational strategies that address the changing landscape of quantum information revolution.

\end{abstract}

\section{Introduction and Framework}
\label{intro}

While transistors, lasers, traditional computers and other inventions continue to enrich our lives, the quantum information revolution is creating excitement across many science and engineering disciplines \cite{Preskill,european,raymer2019,flagship,alexeev2021quantum}. This new frontier of quantum information science and engineering (QISE) is enabled by the exquisite coherent control and manipulation of quantum states of matter using quantum superposition and entanglement that promise transformative improvements in our ability to compute, communicate and sense. The interdisciplinary nature of this revolution spans numerous areas of science and technology \cite{aaronson2013, levymrs, qkd2017, trapped,semiconductorqubit} and has led to rapid advances in quantum computing, communication, and sensing technologies \cite{2020superconducting, circuitqed, kimble, awschalom, photon}. Recent initiatives in the US and Europe, such as the US National Quantum Initiative Act \cite{raymer2019} and the European Commission Flagship kick off \cite{flagship}, in addition to major efforts in China, Japan, Australia, Canada and other countries, have further accelerated progress in QISE. 

The rapid growth in QISE has led to concerns about the growing workforce needs in the field \cite{fox2020cu, singhasfaw2021pt} and preparing students for diverse quantum proficient and quantum adjacent jobs in this interdisciplinary area \cite{meyer2022cu, asfaw2022ieee,muller2023prperworkforce, qtmerzeletal}. For example, in 2021, one conference focused on brainstorming these issues and emphasized that college educators should consider developing courses, curricula and degree programs to prepare students for the fast-growing career opportunities in QISE \cite{asfaw2022ieee}. To meet the QISE workforce needs, it is critical to inspire and educate students in this interdisciplinary field. Therefore, educators must urgently engage students with QISE concepts at appropriate levels via a variety of courses and curricula \cite{kohnle2013, rodriguez2020designing, goorney2024framework, bungum2022quantum, michelini2023research, singh2015review, donhauser2024empirical}. A growing number of educational initiatives have been developed to address these needs, including innovative undergraduate laboratory experiences and instructional tools \cite{beckphoton,singh2007comp,marshman2016ejpphoton,Kohnle_2017,devore2020qkd,maries2020mzidouble,kiko} and creating specialized quantum courses and programs at both secondary and higher education levels \cite{singh2022tpt, qtmerzel, hennig2024new, qtbrang, qthellstern, qtgoorney,qtmeyercu,qtsun,hubloch,hucomputing, lopez2020encrypt, chhabra2023undergraduate, michelini2022, schalkers2024explaining}.

For this work, we use Vygotsky's framework that emphasizes the Zone of Proximal Development (ZPD) and the need for providing guided support and scaffolding to build upon concepts with which students are already familiar \cite{vygotsky}. In the interdisciplinary field of QISE, students come from different disciplines such as chemistry, engineering, computer science, physics and mathematics and have different prior knowledge and preparation as a result of different disciplinary training. For example, only some students may be adept at coding in a computer language such as Python while others may be adept at linear algebra. Ensuring that the QISE courses have appropriate instructional design that embraces students' diverse backgrounds while successfully addressing the challenges of remaining in students' ZPD regardless of their backgrounds is critical. Vygotsky's framework suggests that instructors can transform their QISE classes with students from diverse interdisciplinary backgrounds into communities of practice in which students from different prior preparations have opportunities to learn from each other. These communities of practice can leverage students' varied expertise to enhance learning outcomes for all students while providing tailored instruction that bridges gaps between students' prior knowledge and new quantum concepts and skills to be learned \cite{lave1991situating}. For example, using guided hands-on activities and collaborative learning through creation of a common ground across students \cite{nokes2015better}, educators can enhance learning for students from different disciplines and help them transition from basic quantum concepts to complex applications. Within this framework, engaging in collaborative projects and guided exercises before tackling independent projects can pave the way for students to gradually gain autonomy \cite{shabani2010vygotsky} and help prepare them for a career in QISE.

This paper describes a qualitative research study involving individual interviews with seven university educators who are passionate about QISE education and are also established researchers in this area. These educators have embraced the challenges of developing QISE-related courses and curricula at the undergraduate and graduate levels, and have also advised/mentored students pursuing research in QISE. Their reflections on the challenges and opportunities in developing and teaching these interdisciplinary courses and mentoring students can inspire and aid other educators who are interested in developing their own QISE-related educational programs. Hence, our study focuses on the following main research questions:
 
 \begin{itemize}
     \item[RQ.1] How do educators in QISE design and implement courses and curricula to effectively teach and accommodate students from diverse academic backgrounds, and what types of challenges and opportunities are present in both undergraduate and graduate level courses?

     \item[RQ.2] What insights do these educators provide for advising and mentoring students in QISE and preparing them for research in interdisciplinary lab/research environments?

 \end{itemize}

\section{Methodology}
\label{methodology}

\subsection{Participants and Interview Process}
Eighteen individuals conducting research in quantum-related fields (with implications for QISE) were initially contacted either via email or in person to request an interview. Five declined to participate due to personal reasons. Thus, interviews were conducted with the remaining thirteen educators on various topics related to the quantum information revolution, a portion of which was focused on broadening opportunities \cite{ghimire2025reflections}, countering misinformation in the field \cite{kashyap2025strategies} and current state and future research prospects \cite{liam}. The interviews were conducted online via Zoom, and each interview lasted for 1 hour to 1.5 hours. The interviews were automatically transcribed by Zoom and transcription errors were reviewed and corrected manually by referring to the original interview recordings. In the quotations from educators, we have removed repeated and filler words, such as ``like", ``you know", etc., for clarity.

All the interviewees were leading quantum researchers and had experience teaching quantum-focused courses at the university level. One of the authors knew all but three of the interviewees and approached them about this educational research study. The three educators interviewed, whom we did not know personally, were approached because we knew about their prominent efforts in QISE education. Six of the educators were from the same higher education institution (not all necessarily from the same department) whereas the other seven were recruited from seven different institutions. Three of the educators were faculty members at non-US institutions, however, all institutions are in countries with majority White populations.

The interviews used semi-structured think-aloud protocol \cite{thinkaloud} in which educators were asked many questions that prompted them to reflect upon various issues related to QISE. The style of the interviews was conversational, and the interviewer asked follow-up questions after the educators had discussed their thoughts based upon the initial prompt. Here we only discuss the portion of the interviews that focused on the following questions (with additional follow-ups as noted) related to our research inquiry on QISE education and workforce development:

\begin{itemize}
    
\item[1.]	Considering QISE is an interdisciplinary field, what are your thoughts on how to educate students from different disciplines about QISE? Have you taught any courses in this area with students from different majors? If so, what type of course was it and what considerations did you have to help students with diverse backgrounds? Were you and students satisfied with the course? Any after thoughts? (Educators were also asked about assessment issues in their courses.)

\item[2.]	How do you train students for the work they do in your lab/research group? What are their backgrounds? 

\end{itemize}

Here we focus on the reflections of seven out of the thirteen educators. We used a purposive sampling strategy to select responses which were most aligned with our research questions mentioned in the previous section \cite{patton2002qualitative}. Two main considerations guided our selection: First, we prioritized pedagogical breadth, meaning that we included educators who either described courses and or programs serving students from interdisciplinary backgrounds, or those who explicitly reflected on how to structure such interdisciplinarity in teaching and mentoring. Second, we focused on interviews that provided sufficient details for our analytic approach, particularly in terms of developing courses and curricula and advising in QISE. Following these criteria, six 
interviewees were excluded from our results. Among these educators, half of them exclusively taught quantum mechanics courses to physics major students and did not engage with interdisciplinarity. The remaining educators focused primarily on quantum optics and provided little detail relevant to our research questions. While still valuable, insights from these interviews, which go beyond the scope of our current research questions, will be analyzed in separate research work.

Overall, our final sample of seven educators discussed their reflections on teaching undergraduate/(post)graduate level courses broadly focused on QISE, and their detailed reflections can be very valuable for other educators interested in interdisciplinary QISE courses and curriculum development. Out of these seven educators, four were in physics departments (whom we refer to as educators PH1 through PH4), two were in electrical engineering departments (educators EE1 and EE2) and one was in school of computing and information ({educator CI1}). All seven educators had their Ph.D. degrees in physics or applied physics, even though they now serve as faculty members in other departments. They were all tenure-track faculty at different stages in their academic career, ranging from assistant professor to full professor.

Four educators were experimentalists, and three were theorists. These educators were established researchers and were involved in research focusing on superconducting qubits, spin qubits, NV center qubits, photonics and quantum spectroscopy, and quantum networks and communication. The following specifies the research specializations of each of the seven educators: Educator EE1 conducts experimental research in quantum communications and computation. Educator EE2 conducts experimental research in quantum information using semi-conducting qubits. Educator PH1 conducts experimental research in quantum computing with superconducting qubits. Educator PH2 conducts experimental research in quantum sensing and computing. Educator PH3 is a theoretical physicist doing research in quantum optics with a focus on quantum technologies. Educator PH4 conducts theoretical research in topological and super-conducting  quantum computing. Educator CI1 conducts theoretical research in photonics quantum computation and quantum communication networks.

All seven educators (EE1, EE2, PH1, PH2, PH3, PH4, CI1) discussed undergraduate level QISE courses they had already taught, were planning to teach soon and/or undergraduate level QISE degree programs they had developed. The undergraduate courses and curricula ranged from mandatory for undergraduates, who were in electrical engineering on a quantum computing track, to elective courses for undergraduates across disciplines interested in QISE. 

Three out of the seven educators (PH1, EE1 and CI1) discussed graduate level (this word is synonymous with post-graduate level in some countries) courses they had developed and taught in QISE. Although three graduate level QISE elective courses (not mandatory courses) were taught by educators from three different departments, since students enrolled were from different science and engineering disciplines, the instructors had similar concerns about how to effectively help all students with diverse backgrounds learn in those courses, which are discussed in the results section.

\subsection{Educational Context}
Since five of the seven educators discussed in this paper were in US universities, we provide some context to better understand their reflections for the QISE undergraduate and graduate courses and curricula they discussed. We acknowledge that the context of our study is predominantly grounded in the US setting, despite observing no qualitative differences between reflections of educators from US vs. non-US institutions.

In the US universities, majoring in a discipline as an undergraduate is similar to obtaining an undergraduate degree in that discipline. However, some departments, such as physics and astronomy, may offer several majors within the same department,  such as physics, physics and astronomy, physics and education, astronomy, quantum computing, etc., each with somewhat different course requirements (although all students are typically required to take certain core courses). 
Also, in the US universities, the typical undergraduate course requirement is supposed to take 4 years to complete for a traditional student. 

While some of the courses in the major are considered mandatory courses, others are electives that students can choose from based upon their interest. The courses are typically for three credits with each credit corresponding to a 50-minute instructor contact hour per week (so that a 3 credit course has either three 50-minute contact hours or two 1 hour 15 minutes contact hours) per week. In one semester, full-time students typically take courses totaling 12-18 credits. Occasionally, courses can carry 4 credits. Seminar courses or research courses (in which students can work on research with a faculty member) can typically range from 1-4 credits per course per term. Some students may double major, and some may add certificates in a particular area, e.g., quantum computing and quantum information, which may require 15-20 credits with some courses being mandatory and others being elective for the certificate. 

For the graduate courses in the US, departments such as physics and chemistry typically offer admission for integrated master's and Ph.D. degrees (most students are accepted only if they want to pursue a Ph.D. with full tuition covered as well as a monthly stipend for living expenses) with several courses as mandatory core courses that students must complete in the first two years while others are electives for students to choose from based upon their interest. However, most engineering schools and schools of computing and information in the US admit students separately for master’s and Ph.D. programs with a majority of students (with interest, e.g., in engineering or computer science) only pursuing a master’s degree who pay tuition (unlike the integrated master's/Ph.D. admissions in the science graduate programs focusing primarily on the Ph.D. with a master's degree awarded along the way when a certain level of course work has been completed). The engineering or computer science master's programs involve students completing mandatory and elective courses, some of which could involve research experience with faculty members in their research labs.

\subsection{Analysis}

We used structural coding to organize the qualitative data according to our research questions \cite{saldana2021coding, hedlund2013overview}. The authors discussed several approaches for organizing the data and agreed upon using structural coding as an appropriate method to present the data considering the semi-structured format of the interviews and different follow-up questions. Since structural coding method is a holistic coding approach, using larger segments of the interview transcripts allowed us to identify common themes across the educators' reflections within three broad categories: developing and teaching undergraduate level courses and curricula, developing and teaching graduate level courses and curricula, and strategies for advising and mentoring students pursuing QISE research in their individual research groups. Common themes across the first two categories included course and program development, content and pedagogical approach, interdisciplinarity and diverse backgrounds, student engagement and assessment, and career preparation and evolving landscapes. For advising and mentoring, recurring themes included research training and collaboration, supporting student success and career development, external opportunities and hands-on experience. The authors decided to present the findings in an educator-centered format within each category in order to preserve the authentic narrative flow of each individual's perspective and maintain a holistic view of the findings. Therefore, we primarily present educators' reflections in their own words rather than paraphrasing them to preserve the originally intended meaning to help other educators interested in using similar approaches in their QISE course design. At the same time, to support thematic coherence for readers who are interested in a thematic organization, we provided summary tables with common themes and recurring codes which allows readers to cross-reference across educators and content. This dual-structure can help educators interested in implementing these findings in their own QISE-related courses by referring to portions of the text that is specifically of interest to them. Some cross-referenced subsections align with multiple themes, but we chose the over-arching theme based on the core content.

In the following sections, we will discuss each educator's reflections on undergraduate and graduate courses and programs related to QISE, followed by their reflections on advising and mentoring students.

\section{Reflections of QISE educators about undergraduate courses}
\label{sec:undergrad}

\begin{figure}
    \centering
    \includegraphics[width=\linewidth]{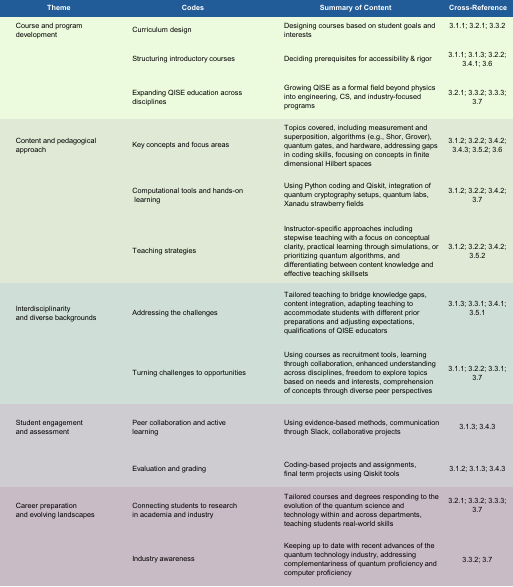}
    \caption{Summary of common themes in educators’ reflections on undergraduate QISE courses.}
    \label{fig:undergrad}
\end{figure}

We first discuss the reflections of the quantum educators pertaining to the undergraduate level QISE courses and curricula they developed, taught, or were planning to teach. Figure \ref{fig:undergrad} summarizes key themes that emerged from educators’ reflections on undergraduate QISE instruction, focused on course and program development, pedagogy, interdisciplinarity, assessment, and career preparation. Each theme is linked to specific codes such as curriculum design, teaching strategies, or peer collaboration and active learning, and is accompanied by a brief summary of the ideas discussed under each code. The figure also provides cross-references to sections of the paper where these ideas are explored in more detail by each educator to provide additional context.

\subsection{Educator EE1's reflections}

\subsubsection{Course and program development}

Reflecting on how the QISE courses she developed are an opportunity for recruiting students for her research lab, Educator EE1 said, “I designed a couple of courses [one undergraduate and one graduate level] that are related to quantum information technologies, and I ask my students [who do research with me] to take those courses. Also, when I recruit undergrads, I usually recruit them only after taking those courses because it's quite a complex phenomenology. I use these courses to see whether students are passionate about the [QISE] topics, because it does require a lot of thinking in different fields and different approaches.” 

Educator EE1 further noted, ``... [the undergraduate course] is aimed at sophomore, junior level students who have taken linear algebra, and then we teach them quantum information and quantum algorithms. From there, if I see that they get excited about this new way of representing and processing information, then that saves us as a group a lot of time, in terms of motivating [students on] why this is interesting to start with." She said that many undergraduates also take the graduate level QISE course she teaches in the following term if they want to pursue these topics further: ``usually students [who] are more advanced keep asking in office hours, but how is CNOT [controlled not gate] implemented? And then I tell them to come to the next class: the graduate level class."

Educator EE1 also noted that the undergraduate course is capped at 40 students, even though they have more students who are initially interested in enrollment. However, after the first week of classes, the number usually settles down to about 30 students. She also mentioned that she makes sure that the homework assigned during the first week is representative of the amount of effort expected from the students, so that they can gauge the rigor of the course correctly and make an informed decision about whether they want to remain in the course or drop it.

\subsubsection{Content and pedagogical approach}

Educator EE1 further reflected on this undergraduate QISE course, describing the content saying, “We try to stay within vectors and matrices, and I do spend probably one class talking about what all these qubits [hardware/architecture] could be. Also in the end [of the undergraduate course] usually…I'll give them one lecture on one technology [hardware] where I ask them [before lecturing on one specific hardware], which one do they want…[Also] I usually encourage them to come to either our seminars or to take the graduate course, because it would be kind of unfair to test them on all of this [hardware]."

Reflecting further on the content, Educator EE1 said, “We need to go through one and multi qubits, quantum gates, DiVincenzo's criteria \cite{divincenzo}, all these things…So in the undergrad course, introduction takes longer and then [toward the end of the term] each class, they do a different algorithm until we run out of classes...we start with Deutsch–Jozsa \cite{deutsch} and go to Shor’s \cite{Shor1994}. There are several [algorithms we do] in between."

Educator EE1 added, ``I don't think 100\% of students understand everything [about Shor’s algorithm]. So the minimum of what I'm shooting for is to understand what are those modules to making this [Shor’s algorithm] happen, not many students are familiar with the Fourier transform to start with. They will pick it up in other courses in all of these departments maybe later, so I do not insist that they need to understand Shor’s from beginning to end. But I do for the Deutsch–Jozsa [algorithm], for example.” 

Educator EE1 elaborated on a component of her assessments mentioning, “analysis of how the wave function is changing in these earlier algorithms as you progress, what happens if we have different inputs…Shor’s algorithm, that's only tested through homework.”

Regarding the textbook used for the undergraduate course, she said, ``For the undergrad course, I used Dancing with Qubits by Robert Sutor from IBM \cite{dancing}. It's a pretty accessible book. I think sometimes there is lack of rigor in the proofs, but that's a good problem to have, because at least the engineers don't complain about not being able to read the textbook." She also referred to a book by Tom Wong \cite{wong} as a new exciting resource that covers a wide range of topics, and said that she may consider switching to it in the future.

Educator EE1 also thought that QISE courses should be different from quantum mechanics: ``I don't think we should be talking about [how] these small particles are wave and particle at the same time...people have learned about this for decades...[what] we have not learned [is] that you can represent quantum information as a superposition of 0 and 1 state; coming down to just a two-dimensional system is what's new, it's more palatable, I think, because when you talk about 0 and 1, it's easier than an integral of all possible states of position or momentum."

Educator EE1 also noted that her students take on coding projects related to Qiskit \cite{qiskit} (requiring coding in Python) which they have developed in collaboration with IBM, and that her students enjoy Qiskit  activities. However, she plans to use a hands-on cryptography setup for her undergraduate course in addition to Qiskit based upon her previous positive experiences in her graduate course. She noted: ``Last year we also introduced the hands-on quantum cryptography setup and that has really improved the retention of the material."

She felt that undergraduates would also benefit from hands-on activities, saying, ``It turns out that if they're using Qiskit, even if it's on real hardware, it's still kind of abstract. And there is something [concrete] when they are rotating polarization or [using] a laser, it just clicks differently than when it's just through keyboard interface.”

Educator EE1 elaborated on the quantum key distribution (QKD) setup she engages students with that uses the BB84 protocol \cite{bennett1984} saying, “This is a very cost-effective system. It's all from the Thor labs for a bit less than \$4,000 and it's called the quantum cryptography analogy setup...It's a pulsed laser and it's not single photon detectors. It's just a detector, but they set it up so that after the beam splitter you have 2 detectors, and you can set them up so that only one shows that you detected randomly; so it kind of simulates as if it was single photon. And students, for demonstration, don't necessarily care whether this is real single photon or not. They're getting the experience of really using the protocol to distribute the key between 2 parties and then there's also eavesdropper. So that's the next level in the demonstration. Then they see how the key gets scrambled through that information, and whether they're able to detect by comparing their parts of key, whether someone tried to tamper with the channel.”

\subsubsection{Interdisciplinarity and diverse backgrounds}

Reflecting on the interdisciplinarity of QISE, Educator EE1 said that she has turned this challenge into an opportunity, “It's something I thought a lot about while I was developing this [course]. I thought that having all these students from different departments is an enrichment opportunity for the course. I use evidence-based methods to make active learning activities in class where students from different departments will work on something together, and already in the first year I taught, I could see that physicists were much more comfortable with quantum mechanics, and engineers were much faster with Python programming, which is good, right? So they could learn a lot from one another. We have a Slack page as well where they're very encouraged to talk to each other for homework. So other than the midterm, everything is very highly collaborative, which is the skill I also hope to impart to them…that's why we kept the undergrad course prerequisite [to be] only linear algebra."

She said that she thinks carefully about what prior knowledge is relevant to connect with students from different backgrounds, for example, when talking about manipulating qubits, ``...during the lecture, I would say for those from physics, you may think about it as a Hamiltonian, or sometimes I would say, for those from engineering, this is an LC oscillator. I would make such comments so that people who know it can make a relationship, and people who don't know it, feel comfortable that they don't have that part of the picture, but that there is a textbook they can read if they want to know the rest of the story. That it doesn't end! That one course in quantum computing is not enough to cover the whole area, I point out throughout. Then [for] the final projects, they need to have at least 2 departments among the 3 students [in each group] and that also helps with understanding material [having people from different backgrounds]."

\subsection{Educator EE2's reflections}

\subsubsection{Course and program development}

Educator EE2 discussed the quantum engineering bachelor’s program in their engineering school. He explained that many undergraduate quantum engineering programs are essentially physics degrees with some quantum information theory integrated. However, they have created a quantum engineering degree program and courses specifically for engineering undergraduates to prepare them to become quantum engineers \cite{quantum_eng_bach_program}. He also noted that one-third of the courses that the undergraduate quantum engineering degree students take are related to quantum and the other two-thirds are traditional electrical engineering courses, saying, ``They have to give up something. They will typically miss out on power systems, energy systems... [but] they will have done control [systems], signal processing, microelectronics very well, probably better than anyone else so they are fully formed electrical engineers [with an emphasis in QISE]. They do all the design courses, [and have] all the real core engineering knowledge.”

Educator EE2 added that the quantum engineering degree program started recently
in response to the evolution of the quantum technology industry, and it also includes a quantum communication course, which is very popular. He also said that “if you want to actually partake of the creation of something new [related to QISE], then this is time to go in.”

\subsubsection{Content and pedagogical approach}
Educator EE2 also noted that they divided an existing quantum engineering course into two separate courses with ``laboratories added" which essentially makes these courses quantum mechanics for engineers \cite{sewani2020coherent, yang2022observing}. He said, ``This starts to look a bit more like a full-on quantum mechanics course but with a lot of Python coding to simulate quantum systems, simulate entanglement and all that." He added that students typically take the first course in their second year and the second one in their third year. He also mentioned that there are approximately 50 students pursuing the quantum engineering degree and some students do double degrees, e.g., finance and quantum engineering, or computer science and quantum engineering.

Educator EE2 also reflected on the fact that in the first undergraduate quantum engineering course, they have minimal expectations from students in terms of prior knowledge, except for some linear algebra. He said, ``The only thing you need is some linear algebra. But…it's only when you see it in practice that you understand the meaning of the algebra. So another thing that happens is that once students take this course, their own understanding of the mathematics and the meaning of linear algebra improves, right? It's all very well to say, what's an eigenvalue and an eigenvector, [but] once you actually see what it means in the physical reality, [it] takes a whole different value.” 

He noted that in the courses he teaches, he shows “how to do one qubit gates, two qubit gates, how to quantify the entanglement in them" but does not go into quantum algorithms and is rather focused on coding and simulations, ``I show the physical basis of entangling quantum operations. I show them how to code it in Python. I show them how to simulate it, how to see what the spins do when they get entangled, how to represent it, how to quantify the entanglement.”

Educator EE2 said that in their first quantum course, they focus, e.g., on concrete systems like quantum dots, “The beauty of using these engineered quantum systems is that because they are finite dimensional systems, it's very easy to write computer codes to simulate them. So that's always been the key part of my teaching, that instead of inflicting a whole semester of partial differential equations on students…we just teach them how to write…Python codes to simulate the dynamics of quantum systems of increasing complexity. You can go fairly complex with almost no additional difficulty in the code. My favorite assignment to give them is to calculate the precession of an electron spin coupled to an increasing number of nuclear spins. The code itself is 5 lines of Python code. There's nothing complicated there. But every time you add that nuclear spin, you double the size of the matrix, and that really teaches the students the complexity of quantum systems. Once you get to 11 or 12 spins, your brand-new laptop with the octa core and everything just becomes a hair dryer and crashes, because you ran out of memory. So that's very pedagogical! And it's very powerful! Because these are engineering students, they know how to code and with very little effort they can learn, really almost hands-on, the behavior of quantum systems.” 

Educator EE2 added, “There are many pedagogical platforms nowadays where you can simulate things but what I insist on is that students write the code themselves. They start from scratch. They write every line of the code. They need to understand how to go from the theory to the application via their own code.”

Regarding the second course focusing on quantum devices and computing, he said, ``it is an expanded version which includes all of the quantum dots, superconducting qubits, cavities, a bit of quantum optics and a bit of quantum information, and it has physical labs. We have actually built…an experiment on nitrogen vacancy centers in diamond. So the students actually learn how to control and read out spins in diamond and do quantum operations, do magnetic resonance on single spins; and the whole…[experiment] is a cheap setup. It’s about \$20,000 worth of stuff, including the commercially-bought lasers and microwave generators which can be repurposed for other projects when not used for this specific teaching lab.”

 Reflecting on the feedback he received, he noted that “I had a good fraction of students who were actually double degree [with] physics, so they had already done two full courses in quantum mechanics under the physics degree and almost invariably at the end of my course, they would come up to me and say, thank you, professor. I now understand quantum physics because that way of teaching by examples and by coding and simulating quantum systems, it's so much more insightful than two semesters of solving partial differential equations.” 

\subsection{Educator PH1's reflections}
 
\subsubsection{Interdisciplinarity and diverse backgrounds}

Educator PH1 reflected on a fundamentals of nanoscience course for STEM undergraduates he has taught a couple of times, which according to him was effectively introductory quantum for science and engineering majors. He mentioned that in this course there are very few pre-requisites and students come from different disciplines, making it a good opportunity to tailor the course to their needs, “I had senior civil engineers and sophomore physicists, and junior biologists. I think in those courses…I just try to figure out what they [want]. I didn't have a really full plan worked out. But when I met them, we did a survey. I asked them what they're interested in…What is the thing you most want to learn? What are you most afraid of? And then I tried to adapt the course to what they're interested in. A lot of them really wanted to learn quantum mechanics and so what we ended up doing was the friendliest version I could cook up of an intro quantum course…and being me, I pick lots of QIS [quantum information science] examples." He also mentioned that since this course does not have to feed into another follow-on course, students are driven by their curiosity and the instructor has the chance to respond to that. ``That's one of the nice things about these [courses], not like sometimes when you're teaching these very structured courses in these long lines of courses, you have to be very concerned with the prerequisite and the post-reqs, and how your [course] stitches into this long path. You have these shorter path courses that are definitely for people who are in there for their curiosity. I would just take as much advantage of that as I could.”

\subsubsection{Course and program development}

Educator PH1 also reflected on the undergraduate Quantum Computing major he recently developed which is relatively new and no undergraduates have completed the major so far. He said that the major combines courses in physics and computer science, and is designed to appeal to two types of students. The first type of students includes those more broadly interested in physics (such as those interested in astronomy, high energy, or condensed matter) who are not necessarily primarily interested in quantum but rather in ``writing complex computer codes and interfacing with instruments through complex coding" which is broadly prevalent in physics. He elaborated, ``Even the theorists are way more likely to be using some big machine to do their calculations than pen and paper. So one part of it [developing a quantum computing major] is that I think every physicist should be more computer proficient. And then hopefully, they go on to be quantum proficient in some quantum industry as well." He mentioned that while the majority of physicists do not take many computer science courses to obtain practical computer knowledge or coding skills, ``the surveys that they do [of physics undergraduates] after their bachelor's [degree] say that 95\% of them take computer-heavy jobs. So I think this [Quantum Computing] major will serve that core need very well." He considered the second type of students to be those who are deeply interested in quantum, saying, ``[for them] the hope is that they are quantum proficient, and that we give them a little different background so that they can talk more to people who are not just physicists; because I see this need for bringing teams of more and more backgrounds together, and working jointly to optimize these really hard [quantum] systems.”

\subsubsection{Career preparation and evolving landscapes}
Educator PH1 also commented on the fact that QISE education efforts are currently dominated by physics, similar to physics’ domination in QISE research, but speculated that the landscape could change over time. Reflecting on why many of the QISE education efforts are currently being led by physicists he took his own department as an example, saying that they are the only department on campus that offers quantum mechanics as part of the regular graduate and undergraduate curriculum. He continued: ``So it's hard to imagine that you're going to be an engineering background student and go through this or an engineering department that hosts this [type of QISE curriculum] if you just don't even have quantum courses. But obviously that is something that can rapidly change. For instance, with transistors, a lot of the silicon physics sort of moved out of physics into engineering and kind of became its own discipline.”

Educator PH1 expressed optimism about broadening of the QISE courses and curricula and felt that QISE courses being offered by departments other than physics would help grow the field, saying, “If the other people don't have educational backgrounds that give them any preparation [in QISE], then it's hard to imagine that we will grow that way. But now you see people in other departments doing this stuff. For now they're still going to take physics courses, but you can imagine that if the field continues to grow, then they'll start teaching their own courses." He felt that departments other than physics teaching QISE-related courses could attract a larger number of students: ``For instance, I think in the engineering school, they're going to offer quantum for non-physicists as an intro course to help people make this transition... If those [QISE] courses start to grow, then it's going to naturally move to other departments…”.

\subsection{Educator PH2's reflections}

\subsubsection{Interdisciplinarity and diverse backgrounds}

Educator PH2 had taught an undergraduate foundations course in quantum computing and quantum information, which is the only mandatory course for a certificate in QISE open to all science and engineering undergraduates with the only pre-requisites being a C grade in calculus 1 and 2 (to ensure that the students have adequate math background for learning relevant linear algebra in the QISE foundations course). Discussing the challenges, Educator PH2 said, ``Of course, the challenging part was that…it's very interdisciplinary and sort of requires a background in mathematics, computer science, physics…So one of the things that I realized as I went through the course was that we probably need a little bit more linear algebra." He mentioned that students tended to lack a deep understanding of linear algebra, which he tried to address in class and through homework assignments. He continued, ``But no matter how many times I do the linear algebra on the board and give them homework and so on, it'll still maybe not quite be enough to really get them the depth, the better understanding of the linear algebra…". Even so, he mentioned that students generally could still get a good understanding of superposition entanglement concepts and how qubits work.

\subsubsection{Content and pedagogical approach}

Reflecting upon the QISE course and his priorities in teaching, Educator PH2 said, “The first thing I wanted to do when I approached the course was to make sure that at least the students will come in and [when] they leave, they have a good understanding of what a quantum computer is, what it does, how it works, and also understand the limitations, because there's a lot of hype in the media and in general…vague statements like, quantum computing will revolutionize climate change…I'm not saying that there are not some seeds of truth in those things, but we are a long way away from something that would actually do something [useful]…so that was…one thing I wanted to accomplish. And I think I did that pretty well, because from both the feedback I got from the students and from their assignments, it seemed like they all understood a large part of that."

Educator PH2 also discussed the algorithms included in the course saying, ``We covered all the algorithms. Shor's algorithm I left [but] I covered every component of Shor’s algorithm, for example, phase estimation, quantum Fourier transform... all that was covered in class including the code examples... we covered obviously, the basic algorithms, Deutsch-Jozsa \cite{deutsch}, Simon \cite{simon}, Bernstein–Vazirani \cite{vaziranib} algorithms...also Grover's \cite{grover} algorithm...and quantum Fourier transform, phase estimation. We went a fair way in terms of algorithms…So they were good with all of those…"

He mentioned entanglement as a topic on which he wished he could spend more time but could not due to being more focused on quantum programming, as he ``wanted them to come away from the class knowing that they can access a [quantum] computer in a certain way and write quantum codes", for which they used IBM Qiskit. However, he noticed a lack of sufficient coding experience in their prior preparation. He mentioned that when students were asked about any prior programming experience, many of them raised their hands. However, doing programming in Python for the class, he realized that there was still a gap. He noted, ``Students said, yes, we have had some Python [experience], but perhaps really not enough, or don't remember enough to actually work with, say, the IBM Qiskit simulations, quantum simulations and things like that...". He added, ``I might have to prepare them better [in the future]. I might have to give them more Python [coding], or maybe…there should be a requirement to have some basic Python introduced in a way that…students should have the ability to take some data and plot it and manipulate it..."

\subsubsection{Student engagement and assessment}

In terms of student engagement, Educator PH2 described his students as eager, curious, and coming up with great questions. As an example, he said, ``They were all fascinated by [and] I spent a fair fraction of the class explaining things like superposition and measurement, and the whole idea of how you do X measurement and Z measurement, and then you do the next measurement again. And now, suddenly, you're getting back something else. So that part, I think, really blew their minds in some way, and I got a lot of questions on that." He said that physics major students tended to be much more eager to learn quantum concepts deeply, asking questions like ``how do you know it's a measurement?". He continued, ``Well, they haven't yet had quantum mechanics so obviously for them this is all completely new material. And maybe the engineers are a bit more focused on applications.”

Educator PH2 reflected further on the undergraduate QISE course and the term projects which involved coding, stating, “I wanted to have them do a term project [involving Python coding using Qiskit tools]. The term project is a little bit more advanced...It gets harder and harder if you just restrict yourself to the drag and drop approach [to make quantum circuits in Qiskit \cite{qiskit}]. So they needed a little bit of [Python] programming. It wasn't much. I think it was some knowing and understanding of how to write a [code] for loops and functions in Python and things like that. But again, even though they said they knew some Python, I think it was still very rudimentary, and I think I'll need to spend more time on that next time to give them a little bit more help along that way.” He addressed the content for the term project further, saying, “...Shor’s algorithm, I left [it] as a term paper. The term projects were not [asking students to] explore a brand-new area completely on their own, that would be too much for them. I tried to give them little pieces from the class that we had taken to a certain extent and then said, okay, go beyond that to the next stage."

Educator PH2 stated that the students had a choice of choosing one particular project to do in groups, and believed that the final projects significantly helped students' learning: ``They worked in groups, so it wasn't just one person working on the project by themselves. I think the project worked out pretty well. I was fairly happy with the final code and reports that they put out...[But] it'd be nice if they could all sample from everyone's projects. So I'm trying to think how to make that a bit more open. I mean, I can post all the GitHub links and things like that, of course.” Educator PH2 explained that there were maximum 3 students in each project group in a class of 30 students, “I opened up a page where they could write their names and the projects that they wanted to work on. I let them use a discussion board to form groups and things like that.”

Overall, Educator PH2 was positive about how the foundations course went, saying, ``The students all said that they got [a lot out of the course] in the reviews. Feedback that I got [hinted] that they learned a tremendous amount, and they appreciated a lot more about nuances of quantum computers and things like that than they did before they came into the class." Still, he expressed a desire to incorporate guest lectures on different qubit architectures in future iterations of the course, saying, ``I think I'll try to work into the next iteration, is try to have maybe some of the professors give 10 minute talks or 5 minute talks on the hardware side of things. I talked about the hardware, but I didn't have time to go into it, and I think they might appreciate a bit more how the actual quantum hardware [work]. Things like that might actually also help them to make things less abstract and more practical. Of course, I showed them pictures [of qubit architectures/hardware], but it’s one thing to do that, and then another thing to have someone who actually works on that hardware to come in and talk to them. Yeah, I might try to do that [next time].”

\subsection{Educator PH3's reflections}

\subsubsection{Interdisciplinarity and diverse backgrounds}

Educator PH3 recalled a rewarding experience he had with teaching quantum concepts to an engineering colleague individually. “One of the more fun experiences I had more than 10 years ago...I started teaching quantum mechanics to one of the engineering professors…he was very committed and came every Friday afternoon essentially for tutorials, but the same [commonly used] quantum mechanics textbooks were very difficult for him to follow, precisely because of all the wave mechanics. What we found was a couple of quantum computing, quantum information textbooks that were based much more on finite dimensional Hilbert spaces; he connected much more easily with that. [He] really started to understand the underpinning concepts of what we were doing, and this led to very fruitful exchanges, where I have since learned quite a bit about flow dynamics. As a result of this I realized also that we have a way of teaching quantum mechanics, which I think is good for physicists but not necessarily the optimal way for quantum information scientists, for people who are not going to need really the strong wave mechanics background.” He added, ``The place where [this consideration] is probably most needed is if we start trying to teach quantum information science courses in a really interdisciplinary way."

Regarding qualifications for teaching QISE courses, Educator PH3 said, ``I don't think it’s a question so much of [does the person] teaching the course have a physics background, it’s more with the person teaching." He mentioned that to improve his own courses ``...I've gone to the computer science colleagues, to the engineering colleagues and said explicitly, ‘What are the things that make no sense to you? How can I explain it to you in a way that will bring things to you?’, and I have adjusted the course in such ways as to do that." He continued, ``Certainly, I could imagine, depending on whether you have the right people to do this, you could imagine quantum information science courses, jointly taught by physicists and engineers, by physicists and computer scientists. It would make perfect sense. I have some very close computer science colleagues at [another] University who would teach a quantum information science course almost identically to how I would teach a quantum information science course. I think that it’s a matter of making sure that we work together with our colleagues from other departments and teach courses appropriately. So it’s not a matter of not having the physicists teach the course, it is certainly a matter of making sure that we have engineers and computer scientists involved in preparing it. I think that you are in a better position once you’ve been interacting with interdisciplinary colleagues in this area. There's no question about that.” He also provided an example of students trusting free IBM quantum information and computing courses because of how the language is accessible to a broader array of people. 

Educator PH3 offered advice for educators teaching interdisciplinary courses in QISE saying, “Tailoring courses to things that make the mathematics more easily accessible while teaching overarching concepts, I think, is quite important. I also think that we need to bring much more of this interdisciplinarity into it. I think it’s coming, anyway. I have great exchanges with colleagues in engineering and in computer science, across different universities, and I think that finding more ways to bridge these fields and to prepare students for an interdisciplinary environment would be very important [for QISE].”

\subsubsection{Content and pedagogical approach}

Educator PH3 provided an example of his experience in terms of implementing a more interdisciplinary approach and noted that he has revamped their fourth-year quantum mechanics course to incorporate some of the previously mentioned ideas. ``I’ve recently reworked our fourth-year quantum mechanics class here at [my university], and part of that was to emphasize again more of the things that we do in finite dimensional Hilbert spaces, more elements of quantum information and entanglement and things like that.” He also emphasized, “But I take out a lot of the more complicated wave mechanics and calculations associated with that, because I think that the concepts are much easier to illustrate often in finite dimensional Hilbert spaces, with a few spins or a few harmonic oscillator levels. And so, I tried to bring that ethos in a little bit into teaching things like perturbation theory always in these finite dimensional vector spaces”.

Educator PH3 drew analogies with teaching algebra-based introductory physics courses and [post]graduate physics courses while emphasizing the need to understand and prepare the course content consistent with the background of students involved. He said, “It’s a little bit like saying that there are a lot of physicists who struggle when they’re asked to teach an algebra-based introduction to physics course, because they have to use a very different language than what they would normally use with their research teams or with physics majors." Educator PH3 saw this as a separate challenge of teaching ``material that you know very well in such a way that the audience is going to take what they need out of it". He elaborated, ``When we’re teaching…graduate courses and things like that, the challenge is making sure that you remember all the details of the physics, that you highlight the right things, and you put them together in such a way that advanced physics students can understand them. The challenge there is 80\% in getting physics content right and 20\% in doing the teaching. Whereas the challenge in teaching the first-year introduction to physics, which is algebra based or to new students...is 95\% in teaching and 5\% in the physics content. I think that we need to see any interdisciplinary course that we teach [e.g., in QISE] in that light. Just because it’s quantum mechanics that we’re teaching shouldn’t stop us from thinking that way. With any form of communication, [it is important to know] who your audience is and how do I present things in such a way that they’re really going to understand. And I think we need more physicists engaging properly with that task [of teaching joint QISE courses].”

\subsection{Educator PH4's reflections: Content and pedagogical approach}

Educator PH4 taught a 14-week long quantum information course that students in any year across various science and engineering disciplines could take. He reflected upon various math and quantum pre-requisites for this course saying, ``I go fully into what's a ket, how to add them up, how to do rotations on them, how to figure out what the measurements [could do], and basis transformations." Educator PH4 considered these as typical standards for a quantum information class, and mentioned covering product space which then allows him to teach entangled vs. not entangled states and how to distinguish them using concrete examples.

He expressed some concern regarding students' knowledge of complex numbers saying, “Everyone has to understand complex numbers [but] I don't know if they [actually] do in this class, because whenever I put complex numbers in [assessments], the success rate of them answering [correctly] goes down or sometimes they make mistakes. It's not like they don't have any idea what it is, but I see a number of mistakes that wouldn't happen if I would use real numbers from time to time. So, I think they know what it [complex number] is, but they're not very strong at manipulating it." He mentioned that rules involving conjugation tend to confuse students.

Regarding knowledge of linear algebra, Educator PH4 said, ``I think the most I ever used [linear algebra] is [for] basis transformation [e.g., transforming from Z basis to X basis]." He mentioned that students can often manage even without a formal understanding of linear algebra: ``There are more things you learn in linear algebra, like eigenvalues, eigenvectors, various decompositions, determinants, which basically I didn't use at all in my class…They do have to know how to multiply matrices, which is technically something you learn in linear algebra. But there's a way around it, you can do these things without actually knowing it, because it's just substitutions." He believed that students often end up using linear algebra in practice without explicitly realizing that they are using it.

Educator PH4 explained that he did not teach about Hamiltonian or eigenvectors and eigenvalues because of the specific way this course was set up, particularly expected by the university system. He believed that ``part of the appeal of that class is that you don't need as many pre-requisites", such as linear algebra or quantum mechanics. He explained, “…I don't actually do Hamiltonians. When I do measurements, it's done in a way where I don't measure the operator. What I do is measure in this basis already. So, it's done in a way where, a lot of stuff that you usually use eigenvalues and eigenvectors for in a more traditional physicsy way, it's just not needed." He believed that the course material was extensive enough that not focusing explicitly on this concept did not create a major gap in students' understanding. However, he believed knowing it would still be useful. ``It would still be nice to have that concept so that [students would know] at certain times, when you apply this operator, this doesn't change [because it is an eigenstate of the operator]. And it shows up in the phase estimation algorithm [which is useful for Shor’s algorithm], it helps to know it ahead of time."

Educator PH4 continued, ``In particular, because I did gate-based quantum computing, it wasn't absolutely necessary to do Hamiltonian...But there were certain times when students would have a question that would be answered if I could tell them what the Hamiltonian was... for example, how do you apply this particular gate…in a physical system, how would this gate work in this case?... for example, you apply some field, you send some pulse, you send some light and stuff like that to kind of explain a little bit about why it [gating] works, you need the Hamiltonian formalism for that.”

Regarding previous knowledge of quantum mechanics, Educator PH4 said, “I think close to half of them [the students] did take quantum mechanics, specifically the quantum I [junior-senior level course] from the physics program, and a bunch of them didn't. So those that took quantum mechanics found the first part [of my quantum information course], which was all about single particle states and ket and superposition, pretty easy because it's almost a repeat from the material. So that's the advantage that those who've already taken quantum mechanics have."  

Educator PH4 noted that algorithms and multi-qubit systems were most challenging for many students. ``People had the most trouble with the algorithm part of the class. Mostly, I think, because I rushed it. It’s new material…Grover, Shor, Deutsch Jozsa [algorithms]… [Shor’s algorithm] wasn't on the test, but Grover's was…I wouldn't expect them to know every detailed aspect of it [Shor’s algorithm], because it is actually kind of complicated, if you were to go through all of it. But I would hope that they understand some of the big picture of why a quantum computer is able to execute Shor's algorithm.” He addressed some changes he would like to make teaching this course again saying, ``I'll probably do entanglement a bit better, because there was a lot of confusion about QKD [using entanglement] versus quantum teleportation. So that aspect I would definitely have to focus more on. But they all did pretty well on just the concepts of single qubits, measurements and transformation and stuff like that. So, I'll certainly still cover some of that [but] next time [will also] do a bit of applications, like quantum sensing.”

\subsection{Educator CI1's reflections: Career preparation and evolving landscapes}
 
Educator CI1 reflected on the growth in QISE and the need to prepare students for the quantum workforce saying, “We see these days [that] lots of students apply for quantum information, and I'm sure it is going to grow, but at the level of the industry and engineering tasks and positions that are gonna come up, I think it's definitely going to be a huge market out there." He mentioned the importance of building larger university programs, saying that ``we in the university, have a great opportunity to service that gap and cater to students."

Responding to how can we help students prepare for the quantum industry, Educator CI1 mentioned, ``Research, a little bit of project experience…if students are involved in any of these big research projects, even from a point of view of an undergrad in their own capacity, whichever way they contribute, that is definitely going to help them." However, he mentioned industry awareness as a critical factor, ``But I think at this point, it's not clear unless we know what kind of positions open up for undergrads in the quantum industry. Once we know that better, we probably can tailor our courses better". He mentioned relevant topics such as basic error correction, quantum algorithms, and quantum sensing, in terms of what he can think of ``in terms of equipping students for the workforce."

Regarding what Educator CI1 would like to emphasize in his undergraduate introductory QISE course that he was planning to teach soon, he said he wants to “give them a sense of what is actually possible, at least as of now, and what is not so, to have a realistic picture of what this whole technology, this revolution, is all about." He continued, ``because right now, there is just a lot of big talk in the industry. A lot of people really don't know what they're talking [about], but they know quantum is a big thing. Everybody knows, oh, AI and quantum are the next big things, but they can't distinguish between what is possible and what is not.” 

Educator CI1 also stated that “since at this moment there is no clear frontrunner in quantum hardware, [I would like] to probably give them a sense of all the different platforms.” He also felt that it would be great “if they have a sense for what would be a breakthrough in each platform that would make things scale up. That would prepare them for the revolution in that platform when and if they are involved in any of those companies or it would make them employable in those particular platforms.” He noted, “I think the core course is a good place for us to teach all…[basics]. They check all these boxes that we spoke about in terms of awareness about the quantum industry, the quantum information arena... "

Educator CI1 also said that the most effective approach would be to develop “courses with the hands-on experience working with the quantum tools." He mentioned that he sees a lot of students who ``are able to pick up quantum everything from scratch by just using these software tools." Therefore, he planned to incorporate software tools in his teaching, noting their success in one-on-one interactions with students, and that he has noticed that students learn from those software tools very quickly when prompted to use them. For example, he described how a student with an electrical engineering background who didn't have a background in quantum optics picked up software tools, ``I introduced her to Xanadu's strawberry fields \cite{xanadu}, and they have a very nice documentation of all the basics of quantum optics. It's not like it can replace a textbook from a pedagogical point of view, but for certain types of students coming from an engineering background, it's so hands-on…I was pleasantly surprised to see that she was already able to take the documentation, take the code and start doing simulations and experimenting with the data in terms of how quantum states behave. There are nice schemes in quantum optics that you can just simulate with these things, and when you simulate you'll learn in a different way than from reading in a textbook. So I see that that's a very powerful way as well."

\section{Reflections of quantum educators about graduate courses}
\label{sec:grad}
\begin{figure}[h]
    \centering
    \includegraphics[width=\linewidth]{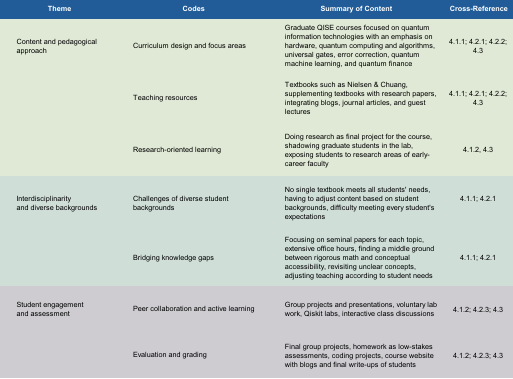}
    \caption{Summary of common themes in educators’ reflections on graduate QISE courses.}
    \label{fig:grad}
\end{figure}

We now describe three quantum educators' reflections on the graduate QISE courses they had taught. Figure \ref{fig:grad} summarizes key themes that emerged from educators’ reflections on graduate QISE instruction, focused on content and pedagogical approach, interdisciplinarity, and student engagement and assessment. Similar to the structure of Figure \ref{fig:undergrad}, each theme here is again linked to specific codes such as curriculum design, bridging knowledge gaps, or peer collaboration and active learning, and is accompanied by a brief summary of the ideas discussed under each code. Figure \ref{fig:grad} also provides cross-references to sections of the paper where these ideas are explored in more detail by each educator to provide additional context.

\subsection{Educator EE1's reflections}

\subsubsection{Content and pedagogical approach}

Educator EE1 taught a graduate course focusing on quantum information technologies that had equilibrated with about 15 students. She mentioned using the textbook by Nielsen and Chuang \cite{nielsen2010quantum}, but she noted, ``because our courses are very interdisciplinary, we are in electrical and computer engineering and our students come also from physics, from computer science, from math, and a few other departments like material science, chemistry, etc., Neilsen and Chuang is really hard to reach. So I keep that only for the graduate course and only use parts of that book." She added, ``For the grad course, that introductory part introduces people to how quantum information is represented, what are bras and kets, what are quantum gates, how does the quantum circuit look like, we cover that in about a week. So we really go quickly, and then we continue with some of the [other] content…"

For this graduate course, she mentioned using a combination of the Nielsen and Chuang textbook with relevant papers in the field, and that the course is ``centered on quantum information technologies and focuses more on hardware, so how different implementations of quantum hardware work." She elaborated that in the course they talk about color centers, superconducting systems, trapped ions, and photonic systems, and that they go through the papers after covering the different hardware to see how they are implemented.

She continued, ``We analyze papers as we go through different technologies and different types of hardware. We look at the very first paper that came for some of those technologies, and then the best, most important paper in the field; and grad students are comfortable with that. They know how to look for information. Undergrads [who are enrolled in the graduate level course] learn a lot through that process because they read a paper, [but] they understand only a part of it; and they haven't often had a course where there's material that they're not supposed to understand 100\%. So we spend quite some time in office hours, explaining that that's okay [not to understand everything in papers], and that's how it looks to enter a research field.”

\subsubsection{Student engagement and assessment}

Educator EE1 explained that students in the graduate course took on final projects, which counted as one form of assessment. The projects were done as group projects with about 3 students per group. She elaborated, ``Students can also opt into doing research in my lab for the final project. So they shadow [current] graduate students [in my research group] and learn to do something, if it's aligning beams or modeling numerically or anything that's at that point the project [my lab] can support. That has also been a really good way for students to try out research and then decide to come and do more.”

She noted that although students could work in her lab to satisfy their course-related project requirement, they often prefer doing a presentation because working in a lab is ``harder than a regular project." She continued, ``Right now, a project would be, say, learn about variational quantum eigensolver [on Qiskit] and apply [it] to something...whereas if they do research with us, it's more involved. They need to coordinate with the lab…”. As an example of how students may perceive doing research as requiring more effort, Educator EE1 mentioned in the last [term] presentation, one student who worked in her lab instead of doing a regular project on Qiskit said, ``It looked easy but I ended up aligning it [laser] for 4 hours...”. She added, ``So it hasn't been too hard that way [to manage final projects] because most groups do something that ends up being a presentation [e.g., what they did with variational quantum eigensolver in Qiskit] and not within my [research] lab [since] we are limited to how many students can do [research with us].” She noted that in both the undergraduate and graduate courses, the Qiskit labs were developed in collaboration with IBM.

Educator EE1 also noted that although these graduate and undergraduate QISE courses are electives, they are trying to see what is the most meaningful way to implement a certificate or diploma, particularly ``because there are students or maybe people returning from industry who want to get expertise in this [QISE]" and she wondered ``what can we do to give them a certificate or diploma that focuses on quantum technologies.”

\subsection{Educator PH1's reflections}

\subsubsection{Interdisciplinarity and diverse backgrounds}

Educator PH1 discussed some of the challenges and opportunities of teaching a graduate-level elective course on quantum information and computing. In this course, there were 14 students from different science and engineering disciplines, and considering their varied backgrounds, educator PH1 stated that teaching such a course is ``more difficult than just doing it for physics major students". However, he found this beneficial for his research and laboratory, saying, ``The more I talk to these other people [students from different backgrounds], the better I see the full picture. I definitely benefit from struggling with these different backgrounds. The course has evolved as I learn more about the other parts that are not just physics; then I can kind of stitch them together in a way that I think makes more sense." Thus, Educator PH1 believed that teaching the graduate level QISE course has made him reflect upon how to interact with and explain core QISE concepts to not only physicists, but also to non-physicists. He believed this benefits his research as well, particularly given his collaborations with non-physicists and the need to effectively communicate with them.

Educator PH1 further reflected on how he optimizes his own goals for student learning in the course with students' backgrounds and expectations saying, “So this idea of joint optimization, I feel like I can start to bring [it] into a course; [knowing that] here the physics says this, the computer science says that. We gotta figure out how to stitch them together. And definitely, this is something I've learned [recently]".

He mentioned using Nielsen and Chuang as a textbook for his course, and regarding the extensive mathematical proofs he said, ``[Considering the different student backgrounds] I'm more confident to say: there's a bunch of math here, here's what he's trying to tell you... and if I can…make it into a story, a more human thing that you can reason about, instead of a bunch of mathematical proofs all layered up, means that everyone who is not a really good mathematician can get more out of it. That helps most people's backgrounds." However, he noted that this means some students' expectations would not be met, ``I think I got one comment in the course saying you didn't do enough math derivation. So there was definitely someone from a heavy math background who was unhappy.” He acknowledged that the very different mathematical preparations of the students in the QISE course was the biggest challenge. He noted that even among physics major students there are differences in mathematical preparations, and elaborated that in the case of students who come from different departments, ``some of these kids, who are very math-heavy, find linear algebra to be trivial [and] some of them find it to be strange and mysterious, so that part [diversity in math preparation] is never going to go away."
 
To address the diversity in mathematical preparation, Educator PH1 stated,  ``You just try to teach to the middle. I don't think you can always teach the lowest common denominator, because then it's too hard to make progress. But you have to understand that you're teaching to the middle, so don't teach to the top. That guy is never going to get his proofs, because most of my students would not hang in there for him. On the other hand, I'm not gonna really do linear algebra a full semester...You try to find the middle ground and hope that people can converge on that.”

Educator PH1 also thought that the fact that the course was a graduate level elective, rather than mandatory, was beneficial. He stated, ``So it's not like I'm trying to get the pre-req[uisites] for the next course that they're going to take, and make sure that the next professor is ready to take them forward. It's really meant as a capstone of your graduate career." He believed that his role was to provide students with tools and knowledge to independently engage with the material later as they needed rather than focusing on specific proofs. He continued, ``I still think I can do a good job if I teach you the shape of the endeavor, and how to go find the pieces that you need later; like how to engage with the literature in a more productive way. So that is one of the nice things about…[an elective], that it's meant for people who are going to engage with it and dwell with it for a while...because they're meant to go do research. They're meant to go push on the boundaries. They're meant to go find their niches. I'm just trying to make it make enough sense globally that they can at least relate to the idea that there is a field here that's trying to do this.”

Educator PH1 said that most people were satisfied with the course and further reflected on how he adapted his graduate course to student needs, “Most people were pretty happy in the end. I had some grumbles...but also [a class with] 14 [students] is not so big, so we were pretty responsive. If students asked a lot of questions about X, the next lecture I could talk about X, or circle back to something that didn't make sense...So I definitely did a lot of Q\&A with the class and adjusted. My plans were based on the things they thought were really interesting or the things they found really confusing.”

\subsubsection{Content and pedagogical approach}
Regarding the content covered, Educator PH1 noted that while he did not cover networking, they did cover computation and the first three chapters of the Nielsen and Chuang textbook as an introduction to QISE. Describing the content further, he said, ``...[we went over] universal gate classes, [and] decomposition. We talked a fair bit about how you build gates out of other gates, [and] how you compile or transpile. That took the majority of the semester to get through.” He emphasized that the first 65\% of the course was an introduction to “what is quantum mechanics for two level systems and also how it turns into computation for gates and universality”. He continued, “Then in the last third of the semester, we talked about error correction, and codes in general, plus I felt like we had to do Shor’s and Grover’s [algorithms], just so that they would have seen two of the top line, two of the most famous algorithms.”

Educator PH1 also reflected on why he felt covering Shor’s and Grover’s algorithms was important in this QISE course and how Shor's algorithm can be too intricate as the order finding algorithm, “The order finding algorithm is really only interesting if you have a background in modular arithmetic and digital math and factoring. Most of my students do not have a background in any of that. It's also a non-deterministic algorithm. It involves phase estimation and doesn't even involve the quantum Fourier transform. It only involves the inverse for the Fourier transform...So I like Grover’s [algorithm], because I can at least explain it....this graphical interpretation I find to be rather beautiful."  
Being deeply involved in research to build a quantum computer, he further added, ``I want to build a thing, and have it exist. [But despite my research interest] I feel I can't skip Shor’s because it's the most famous one.”

\subsubsection{Student engagement and assessment}

Regarding the assessment of the graduate elective course, Educator PH1 reflected, “We did have homework. I graded it, but not very heavily. It was mostly there to give [the students] something concrete to do." This way, he ensured that the students are actively engaging with the material rather than only discussing the concepts in the class. He added, ``We did do projects, and this time I asked them to actually code up their projects. I think 90\% of them did so, for instance, in Qiskit or something like that, and they could either work an algorithm and show that if you put an input, you can get an output, or they could put a code in. I thought that was really nice." He mentioned using class discussions as an opportunity to enhance their understanding of why a code failed, saying, ``Definitely, several of them bit off way more than they can chew, but even though their codes crashed and burned on the day [they presented], just even talking about what they were doing and trying to see where the problem was, they learned a lot." Without using heavy assessments, he provided enough activities for students to engage with the material for this elective course.

\subsection{Educator CI1's reflections: Content and pedagogical approach}
 
Educator CI1 reflected on a QISE graduate elective course he taught in which he had 10 students from the physics department and the school of computing and information. He emphasized that he covered “all aspects of quantum” in this graduate course. To help students learn, he had them read journal articles and made a course website with blogs and the final write-ups of students. He also noted that although this graduate level QISE course was in-person, he had 5 external guest speakers on Zoom join the class to discuss different papers that students had previously read, or the topics they had learned about in the course. This facilitated meaningful interactions between the students and speakers. The guest speakers were early-career faculty and postdocs from other universities. Educator CI1 elaborated, “It was a good opportunity for them to talk about their research. We had one on quantum machine learning, one on quantum finance, one on quantum security…one on quantum sensing, 
who spoke about the dark matter...it's been a hot topic lately with quantum sensing…” Thus, Educator CI1 in his QISE graduate level elective course incorporated topics such as quantum machine learning and quantum finance, which were not mentioned by other interviewed educators from physics and electrical engineering departments. His philosophy on graduate courses was that while these courses can never fully prepare students for research in advance, they can help them stay open to learning, saying ``I don't feel that for a Ph.D. you can ever be fully prepared with all the basics [from courses]. You pick up skills and tools and theories along the way as you go, as many of us have done. I think students just have to keep an open mind and be willing to learn new areas."

\section{Reflections on advising and mentoring students in research}
\label{sec:advising}
\begin{figure}
    \centering
    \includegraphics[width=\linewidth]{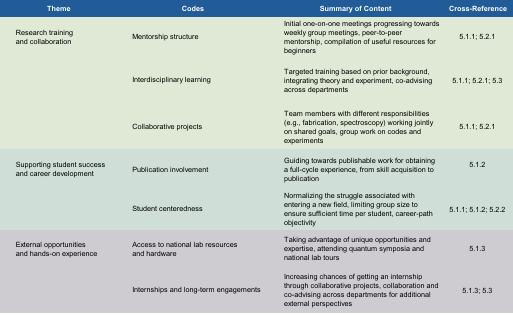}
    \caption{Summary of common themes in educators’ reflections on advising and mentoring students in QISE.}
    \label{fig:advising}
\end{figure}

In this section, we discuss reflections of quantum educators about advising and mentoring, focusing on insights that we felt can be helpful for others in advising and mentoring students in this interdisciplinary QISE field. Figure \ref{fig:advising} provides a summary of the themes and codes in this section. Themes in this section include research training and collaboration, supporting student success and career development, external opportunities and hands-on experience. Use of collaborative projects, involvement in publications, and access to national lab resources were among the discussed ideas. Summary of the content falling under each theme, as well as cross-references to provide additional context by educators, are also provided in Figure \ref{fig:advising}.

\subsection{Educator EE1's reflections}

\subsubsection{Research training and collaboration}

Educator EE1 reflected on how she recruits students in her research group, saying, “We look for students that have some relevant backgrounds. Clearly, no one will have all of that background [in the interdisciplinary field of QISE]." However, she provided an example of how the training usually goes. ``If a student is a more typical physics student, who has a background in quantum mechanics, then we train more [in] the engineering side of things, [like] how we design these devices for specific wavelengths, for specific performance, [or] how we fabricate them.” Therefore, training is provided in a way that targets areas where students may need additional support.

Educator EE1 noted that they have weekly group meetings where graduate students mentor undergraduates to help them with the projects. She also added, “Beyond just those meetings, when a student starts in my group, I give them a few papers and just ask them to write down all the words they don't know in these papers, so that we can quickly push through the language barrier. So we try to get it out of the way in the first 3 weeks, for example, so that they at least know what's the language, and then slowly, really start to understand the more complex phenomenology that is typical for the field. In the [research] lab, they usually shadow the more experienced person in the lab.”

Regarding training students, Educator EE1 continued, “I keep those first student meetings one-on-one, so that students can feel free to be vulnerable to share that they've never heard of certain words, and that...they can ask other students for extra support. But in the beginning, I really want to show them that's totally fine, that those words are new, and that they were once new for me as well, and I think that helps."

Educator EE1 mentioned that to get students started on research in her group, she had compiled a list of references for students to start with. She further explained, “It has a bunch of papers that are important for the start of the field, and [covers] some of the first demonstrated methods. Students don't necessarily read all those. I think there are probably about 30 references, they [students in the research lab] see how they connect...[for example,] first this person had this discovery, and then this experiment showed that you could take it to a different level, etc. We focus on a few review articles, or very specific scientific articles that are relevant for the first project the students are gonna work on."

Educator EE1 also emphasized that her research group is “highly collaborative". She explained, ``For example, we have students who work on fabrication, and then students who work more on equipment in the lab or the spectroscopy and then they come together. They develop their methods, and then those who fabricated bring the devices to the spectroscopy lab and they measure together so it {[collaboration]} is necessary." She further explained how it starts with logistics; e.g., understanding what is on the chip, how far are the devices scanning, what color center is in the system, saying, ``All these subgroups have read papers that relate to setting expectations of what we should see in a specific chip, and then they sit together, either measure together, or look together through the data and try to get a bigger picture of what's going on, what went wrong, what can be done to improve it, etc.”

\subsubsection{Supporting student success and career development}
Regarding undergraduates working in her lab, Educator EE1 said, “I don't have a fixed number of undergraduate students, but we usually have 3 to 6 undergraduates actively working with our research group because I want to…invest [my] time into their understanding so that they really get something out of it, and so far it has worked really well. Students have been able to get some nice positions whether it is for grad school or in national labs." She also added that for the students in her lab, 
``we try to have a publication for each [undergraduate] student that works with us...so that they can see the whole process; from learning a new skill to applying it, to that being just at least a small piece of a bigger result.”

\subsubsection{External opportunities and hands-on experience}
Educator EE1 also emphasized providing internships and offering other hands-on experiences for students working with her. Having been recently asked by DOE (USA Department of Energy) about what they can do and having access to two quantum computers, she elaborated, “I have user grants [with two DOE labs], so having such resources, the training [of my students] becomes so important…these resources are not just the hardware. It's really the expertise of people who are trained and our students are really benefiting from working on those projects. This is for both grad and undergrad students." She also emphasized the importance of shared hardware resources such as dilution refrigerators saying, ``Not everyone has that infrastructure at their university."

Educator EE1 also added that the majority of the collaboration with USA national labs is remote, except for summer internship opportunities for students. She explained, ``We have a collaboration project, and every 2 weeks we meet on Zoom and get access to expertise by those teams because they know a lot of things, and they can help us push our project and there they offer internships [in summer to students] at all levels…So having experience on a project [with a national lab throughout the year] can probably increase chances of [students] actually getting the [summer] internship. So one of my students is going there for a month this summer to do hands-on research differently than what we have been doing over Zoom, where we did most of the modeling and the circuits were run by the staff…”. Educator EE1 also noted that she sends her students to attend occasional quantum symposiums or lab tours that the national labs or industries nearby her university hold, which can be a valuable experience for them professionally.

\subsection{Educator EE2's reflections}

\subsubsection{Research training and collaboration}
Educator EE2 reflected on how they have integrated theory and experiment in his research group saying, “We are an experimental group, but we pride ourselves in also having some creativity and theoretical abilities, so I see that my students understand the theory very well, and for the most part they also develop some of the theory themselves...We can actually design the experiments, design the device, design the quantum gates, and do the theory for it before we actually make it in the lab. So people [students and postdocs in my lab] have a much better understanding of what to look for."

Educator EE2 explained that they have no issues when it comes to training younger students to do experiments because their group is fairly large. He clarified what he means by doing an experiment, saying, ``Nowadays doing an experiment means writing Python codes to control the instruments and gather the data. There is almost no manual part to it anymore. The only manual part is mounting the chip in the refrigerator, connecting the cables, connecting the instruments; from there on it, is just a giant Python code. So that's what doing an experiment means today." He added, ``But at the same time, it's the same code that you use to then simulate the experiment. So it's nice that there is almost no disconnect anymore between the theory and the experiment. You know, in the old days you would write [using] pen and paper, solve partial differential equations, and then go in the lab and turn knobs, [but] right now it's all the same. You write Python code to do the theory, and you write Python code to do the experiment, two things that are almost joined at the hip.” Educator EE2 noted that several of his students/postdocs work in small groups of 2 or 3 people with senior students training the junior ones on these kinds of things. However, he added that “The seniority is very flat. Often, I have cases where a second-year Ph.D. student is training a postdoc, just because that student knows that stuff better than the postdoc…It doesn't matter. It is just [that] wherever there is the knowledge, [it] gets passed on across the people in the team.”

\subsubsection{Supporting student success and career development}
Educator EE2 further noted that students in his lab mostly get jobs in industries such as large corporations and startups including quantum computing startups across different continents, “But some are continuing an academic path…I'm an academic at heart so I always like it when my students themselves become academic, but I know that there is a lot of appeal for students to instead work in a company where they feel that they don't have to go through the same hoops and do the same amount of administration and teaching that an academic is perceived as needing to do...But anyway…they can do whatever they want. So my students never have problems getting whatever job they want..."

\subsection{Educator PH1's reflections: Research training and collaboration}

Educator PH1 reflected on the benefits of co-advising students from another department on QISE research saying, “Actually I see at least in my own life, compared to 10 years ago, that I interact with much more people of these kind of different backgrounds to try to figure out how to make a joint step forward…for instance, I co-advise a student who's in ECE [electrical and computer engineering] and we write these papers specifically where he says, from the gate perspective, from the computing perspective, here's a good choice. And we say, from the machining and physics and nanoscience perspective, here's a good choice. Then we try to look at how they meet up and what's the joint good choice, which is not usually the same as either choice by itself. I think there's gonna have to be a lot of that kind of [collaborative] stuff to build machines [quantum computer] that work well enough.” He felt that co-advising a student from another department has not only helped the student think about these interdisciplinary issues critically but has also helped him think deeply about the challenges in QISE research and the necessity to tackle the issues from many different angles.

\section{Discussion and summary}

Helping students at all levels learn QISE concepts through courses and degree programs, and providing opportunities for them to have project-based experiences commensurate with their interest in this interdisciplinary field, can be valuable for inspiring them to become future contributors and leaders in this exciting field. This research focuses on findings from individual interviews that other educators interested in QISE education would find valuable.  In particular, in the qualitative research discussed here, we focused on the reflections of seven university quantum educators, who are established researchers, about how they are preparing students for the quantum information revolution via undergraduate and graduate level courses and curricula they have developed, as well as by advising/mentoring students on individual research in their labs. Their reflections can be invaluable for all educators considering effective approaches to prepare undergraduate and graduate students for this rapidly growing interdisciplinary field.

Our first research question was focused on designing and implementing courses and curricula to effectively teach and accommodate students from various academic backgrounds. Educator reflections for this part were centered around course and program development, content and assessment, challenges and opportunities regarding interdisciplinarity of QISE, and future career perspectives. 
All undergraduate and graduate level QISE courses which the educators talked about had a diverse mix of science and engineering students, with the exception of those that were part of a quantum engineering degree program based in an engineering school. In the latter program, most students (roughly 50 per cohort) were taking two-thirds of their courses similar to a traditional electrical engineering degree and one-third related to QISE. Some of the educators mentioned that they appreciated interacting with students from different departments who were enrolled in their courses, and that their questions and feedback helped them not only polish their approaches to teaching students from diverse backgrounds, but also improve their communication skills to connect better with other QISE researchers with different disciplinary training. Several educators also described the undergraduate and graduate courses as opportunities for recruiting students for research in their labs.

The educators reflected on the diversity in student preparation as a challenge they needed to address, regardless of whether the courses were at the undergraduate or graduate level. For instance, educators noted that students from different science and engineering disciplines had very different mathematical preparation and they, e.g., had different levels of interest in the “quantum” vs. “computing” part of a quantum computing course. In addition to differences in mathematical preparation, students’ level of prior exposure to quantum concepts or Python programming also varied greatly. The reflections of these educators provided additional fine-grained details of how they uniquely addressed some of these previously established challenges, such as the trade-off between course accessibility and stronger prior preparation \cite{meyer2022cu}. Previous findings from surveying instructors in QISE courses have shown that some instructors do not find additional pre-requisites, such as quantum mechanics or even linear algebra, to be necessary for student learning in QISE courses. Our results show that in fact, many of the educators took advantage of the fact that most of these courses were electives with no pre-requisites or post-requisites to tailor their courses and maximize learning for the specific student population present in their courses. Additionally, they considered minimal pre-requisites as an appealing feature that attracts more students to enroll in these courses based on their own interests. Therefore, not having pre-requisites such as quantum mechanics or even linear algebra was not seen as a limitation (despite requiring additional effort to support students); instead, it was mostly regarded as a positive feature that allows students from different levels and disciplines to learn the basics of QISE.

In addressing the challenge of varied backgrounds, our findings indicated that the educators kept an open mind and adapted the level and coverage of their QISE courses to make the learning meaningful for the majority of students. We found multiple instances where educators' reflections directly embodied Vygotsky's Zone of Proximal Development (ZPD) \cite{vygotsky}. More specifically, the educators emphasized leveraging students' existing knowledge as a foundation for coming up with their teaching and advising practices. For example, Educator EE1 mentioned using different analogies tailored to students’ backgrounds (e.g., Hamiltonians for physics majors, LC oscillators for engineers). She also fostered peer collaborations across disciplines, enabling students to support each other to stretch their ZPD appropriately (e.g., physicists assisting peers with quantum concepts, and engineers with coding). Educator EE1 also provided tiered course offerings (undergraduate level followed by graduate level), coupled with hands-on tools like Qiskit and quantum cryptography setups to allow students to first engage with more accessible concepts and then move towards more complex content to remain in their ZPD throughout.

Similarly, Educator EE2 exemplified focus on ZPD through introducing incremental complexity in Python coding examples, allowing students to move from basic theoretical knowledge to simulating the dynamics of quantum systems by themselves to have a better understanding of the behavior of quantum systems. Additionally, Educator PH2 acknowledged that with minimal pre-requisites for the course, students may lack some mathematical and coding skills. Therefore, he responded to these gaps by revisiting concepts in linear algebra and integrating coding exercises to adapt to what students could realistically achieve with guided support. Many of the projects and assignments that the educators mentioned focused on collaboration and group work as a way to bring together students from diverse backgrounds and address differences in their prior preparation. These efforts were aligned with our framework which promotes mutual support in learning and scaffolding to bridge knowledge gaps, and the idea that learning is fostered through forming and participation in communities with a common goal across the members \cite{lave1991situating}.

Regarding the content covered, regardless of the level of the course, all educators focused on basic quantum concepts relevant for two-state systems, and all but two included quantum algorithms as well. Some educators did not focus on specific qubit architectures (hardware) in detail in their undergraduate courses, but did so in the graduate level courses. Educator PH2 noted that in the future iteration of the foundations of quantum information and quantum computing course, he would like to focus more on entanglement and invite guest lecturers who do research on different qubit architectures to his class. Only Educator CI1 included machine learning and quantum finance in his course materials in the graduate course. 

In addition to lectures, pedagogical approaches in both undergraduate and graduate courses included Python coding to some extent, with many of them using Qiskit tools \cite{qiskit}. However, the educators based in electrical engineering departments (EE1 and EE2) incorporated Python coding as a central component of the course throughout the majority of their undergraduate QISE courses. For example, Educator EE2 described how Python coding was integral to student learning throughout the semester in the quantum engineering degree program courses they had developed. Both undergraduate and graduate QISE courses typically included some end-of-the-semester projects (often involving coding on Qiskit and students working in small groups) and presentations in class. In some cases, students could also opt to work in the educator’s research lab instead of doing a coding project.

Educator EE1 incorporated hands-on activities, such as a QKD setup, to supplement her graduate QISE course and planned to incorporate it in her undergraduate QISE course as well. Educator EE2 also discussed hands-on lab experiences in both their foundations of quantum engineering course and quantum devices and computing course (e.g., a lab based on the NV center). In the graduate level QISE courses, educators also incorporated reading and discussions of journal articles. Educator CI1 mentioned inviting early-career researchers to join via Zoom to discuss their research with students, providing the students an opportunity to interact and engage with them. 

Most educators were appreciative that undergraduate and graduate students enrolled in their elective QISE courses driven by curiosity, and they assessed student learning based on traditional homework, coding projects, and presentations, and generally took a flexible approach to grading. Notably, educators from engineering and computer science were more likely than physics educators to emphasize an intention to align their QISE courses to the needs of the quantum industries in which  students would find jobs. Overall, educators tailored their instruction based on specific student needs and backgrounds (e.g., prioritizing practical skills such as coding and simulation), which also reflects various pathways for students to engage with the field.

QISE is an emerging field, and the courses and curricula are in early stages of development. In this study, we did not aim to establish a standardized framework for QISE course development nor did one emerge from our findings, which further reflects the rapidly evolving nature of the field and instructional demands in QISE. Nevertheless, common themes in both content and pedagogy, as previously discussed, were identified. What was made clear based on reflections of educators in this research is that in developing interdisciplinary QISE courses, instructors often prioritize accessibility and adaptability rather than strict standardization. Therefore, flexible course design and iterative pedagogical approaches that are frequently adjusted based on ongoing assessment of student understanding become particularly effective strategies in QISE teaching. Prior research has also argued that the variety in instructional approaches is a strength of the field, and that while some instructional methodologies may be overall considered more effective, there is no single approach or model (including determining pre-requisites) for QISE course and curriculum development
\cite{meyer2022cu}. 

As quantum technologies move toward mainstream adoption, supporting broad students access from introductory levels and preserving flexibility in prior background remains to be necessary. Findings from our study supported this view: some educators intentionally moved away from traditional wave mechanics approaches, indicating that formal knowledge of quantum mechanics is not necessary and some chose not to formally require linear algebra knowledge, but rather integrate it directly as needed into the course while teaching. Our findings have extended prior research by taking a qualitative approach prioritizing depth over breadth, where focusing on each educator's reflections revealed how they reason and adjust their decisions in response to local contexts, such as student background and feedback, as they go through their courses.

Our second research question was focused on advising/mentoring students in QISE and preparing them for interdisciplinary lab or research environments. Educators' insights in this section were centered around research training through mutual collaboration, supporting student success and career development, and providing external opportunities and hands-on experiences such as access to resources and internships. The educators emphasized on understanding students' prior preparations, as well as identifying their strengths and areas that need improvement before they launch into research. Additionally, they suggested that having several students work as a team on a particular project, and having senior students mentoring juniors can also foster a supportive learning environment. Keeping a list of important papers (including review papers) for new group members to read and discuss was also mentioned as an effective approach. Some of the educators also suggested asking students to take some relevant courses in the discipline before or while pursuing QISE research as a way to improve student outcomes in the research lab. Individual meetings with students who are new to the QISE field and may be feeling vulnerable were also suggested to be particularly useful to address their questions and concerns, e.g., about the papers they may have read, especially for students from traditionally underrepresented groups who are more prone to societal pressures and anxiety about discussing doubts in front of their more experienced peers \cite{mariesnature}.

Themes of taking advantage of the interdisciplinarity of QISE also emerged, e.g., when educators spoke about advising students in the field. For example, Educator PH1 emphasized that co-advising a student from electrical and computer engineering has made it clear to him that such interdisciplinary collaborations are critical to making major advances in the field. This is because often researchers with different backgrounds (e.g., physics or engineering) on a given project only focus narrowly on the issues they understand deeply and overlook other major issues outside their research discipline. However, to make progress and optimize, e.g., a particular qubit architecture's usefulness in a particular application, it is necessary to collectively consider all these issues with the views of experts from various disciplines. Educator PH1 also expressed optimism that more QISE courses and curricula will be developed by departments other than physics in the coming years, as more faculty with an interest in QISE are recruited in those departments. 

Given that the quantum information revolution is still in its infancy, we hope that the reflections of these educators will be valuable for others who are also interested in developing QISE-related courses and curricula at various levels within their own institutions.

\section*{Acknowledgements}
We thank the National Science Foundation for award PHY-2309260. We are grateful to all educators who participated in this research. We also thank Educator EE2 for sharing the links for the quantum engineering degree and NV center-based lab to include in this paper.

\section*{Data availability statement}
Interview transcripts supporting this study’s findings are restricted to protect participants' confidentiality. The data cannot be made publicly available upon publication because they contain sensitive personal information. The data that support the findings of this study are available upon reasonable request from the authors.

\section*{Ethical statement}
This research was carried out in accordance with the principles outlined in the University of Pittsburgh Institutional Review Board (IRB) ethical policy, the Declaration of Helsinki, and local statutory requirements. The educators provided consent for use of the interview data for research and publication, and consent for quotes to be used.


\section*{References}
\bibliographystyle{iopart-num}
\bibliography{refs}

\providecommand{\newblock}{}
\begin{thebibliography}{10}
\expandafter\ifx\csname url\endcsname\relax
  \def\url#1{{\tt #1}}\fi
\expandafter\ifx\csname urlprefix\endcsname\relax\def\urlprefix{URL }\fi
\providecommand{\eprint}[2][]{\url{#2}}

\bibitem{Preskill}
Preskill J 2018 Quantum {C}omputing in the {NISQ} era and beyond {\em {Quantum}\/} {\bf 2} 79 ISSN 2521-327X \urlprefix\url{https://doi.org/10.22331/q-2018-08-06-79}

\bibitem{european}
Acín A, Bloch I, Buhrman H, Calarco T, Eichler C, Eisert J, Esteve D, Gisin N, Glaser S~J, Jelezko F, Kuhr S, Lewenstein M, Riedel M~F, Schmidt P~O, Thew R, Wallraff A, Walmsley I and Wilhelm F~K 2018 The quantum technologies roadmap: a european community view {\em New Journal of Physics\/} {\bf 20} 080201 \urlprefix\url{https://dx.doi.org/10.1088/1367-2630/aad1ea}

\bibitem{raymer2019}
Raymer M~G and Monroe C 2019 The {US} national quantum initiative {\em Quantum Science and Technology\/} {\bf 4} 020504 \urlprefix\url{https://doi.org/10.1088/2058-9565/ab0441}

\bibitem{flagship}
Castelvecchi D 2018 Europe shows first cards in€ 1-billion quantum bet {\em Nature\/} {\bf 563} 14--15 \urlprefix\url{https://doi.org/10.1038/d41586-018-07216-0}

\bibitem{alexeev2021quantum}
Alexeev Y, Bacon D, Brown K~R, Calderbank R, Carr L~D, Chong F~T, DeMarco B, Englund D, Farhi E, Fefferman B, Gorshkov A~V, Houck A, Kim J, Kimmel S, Lange M, Lloyd S, Lukin M~D, Maslov D, Maunz P, Monroe C, Preskill J, Roetteler M, Savage M~J and Thompson J 2021 Quantum computer systems for scientific discovery {\em PRX Quantum\/} {\bf 2} 017001 \urlprefix\url{https://link.aps.org/doi/10.1103/PRXQuantum.2.017001}

\bibitem{aaronson2013}
Aaronson S 2013 {\em Quantum Computing Since Democritus\/} (Cambridge University Press) \urlprefix\url{https://doi.org/10.1017/CBO9780511979309}

\bibitem{levymrs}
Eckstein J and Levy J 2013 Materials issues for quantum computation {\em MRS Bulletin\/} {\bf 38} 783--789 \urlprefix\url{https://doi.org/10.1557/mrs.2013.210}

\bibitem{qkd2017}
Liao S~K, Cai W~Q, Liu W~Y, Zhang L, Li Y, Ren J~G, Yin J, Shen Q, Cao Y, Li Z~P and Others 2017 Satellite-to-ground quantum key distribution {\em Nature\/} {\bf 549} 43--47 \urlprefix\url{https://doi.org/10.1038/nature23655}

\bibitem{trapped}
Bruzewicz C~D, Chiaverini J, McConnell R and Sage J~M 2019 Trapped-ion quantum computing: Progress and challenges {\em Applied Physics Reviews\/} {\bf 6} 021314 \urlprefix\url{https://doi.org/10.1063/1.5088164}

\bibitem{semiconductorqubit}
Zhang X, Li H~O, Cao G, Xiao M, Guo G~C and Guo G~P 2019 Semiconductor quantum computation {\em National Science Review\/} {\bf 6} 32--54 \urlprefix\url{https://doi.org/10.1093/nsr/nwy153}

\bibitem{2020superconducting}
Kjaergaard M, Schwartz M~E, Braum{\"u}ller J, Krantz P, Wang J~I~J, Gustavsson S and Oliver W~D 2020 Superconducting qubits: Current state of play {\em Annual Review of Condensed Matter Physics\/} {\bf 11} 369--395 \urlprefix\url{https://doi.org/10.1146/annurev-conmatphys-031119-050605}

\bibitem{circuitqed}
Blais A, Girvin S~M and Oliver W~D 2020 Quantum information processing and quantum optics with circuit quantum electrodynamics {\em Nature Physics\/} {\bf 16} 247--256 \urlprefix\url{https://doi.org/10.1038/s41567-020-0806-z}

\bibitem{kimble}
Kimble H~J 2008 The quantum internet {\em Nature\/} {\bf 453} 1023--1030 \urlprefix\url{https://doi.org/10.1038/nature07127}

\bibitem{awschalom}
Awschalom D, Berggren K~K, Bernien H, Bhave S, Carr L~D, Davids P, Economou S~E, Englund D, Faraon A, Fejer M {\em et~al.\/} 2021 Development of quantum interconnects (quics) for next-generation information technologies {\em Prx Quantum\/} {\bf 2} 017002 \urlprefix\url{https://doi.org/10.1103/PRXQuantum.2.017002}

\bibitem{photon}
Zhong H~S, Wang H, Deng Y~H, Chen M~C, Peng L~C, Luo Y~H, Qin J, Wu D, Ding X, Hu Y and Others 2020 Quantum computational advantage using photons {\em Science\/} {\bf 370} 1460--1463 \urlprefix\url{https://doi.org/10.1126/science.abe8770}

\bibitem{fox2020cu}
Fox M, Zwickl B~M and Lewandowski H 2020 Preparing for the quantum revolution: What is the role of higher education? {\em Physical Review Physics Education Research\/} {\bf 16} 020131 \urlprefix\url{https://link.aps.org/doi/10.1103/PhysRevPhysEducRes.16.020131}

\bibitem{singhasfaw2021pt}
Singh C, Asfaw A and Levy J 2021 Preparing students to be leaders of the quantum information revolution {\em Physics Today\/} \urlprefix\url{https://doi.org/10.1063/PT.6.5.20210927a}

\bibitem{meyer2022cu}
Meyer J~C, Passante G, Pollock S~J and Wilcox B~R 2022 Today's interdisciplinary quantum information classroom: Themes from a survey of quantum information science instructors {\em Physical Review Physics Education Research\/} {\bf 18} 010150 \urlprefix\url{https://link.aps.org/doi/10.1103/PhysRevPhysEducRes.18.010150}

\bibitem{asfaw2022ieee}
Asfaw A, Blais A, Brown K~R, Candelaria J, Cantwell C, Carr L~D, Combes J, Debroy D~M, Donohue J~M, Economou S~E {\em et~al.\/} 2022 Building a quantum engineering undergraduate program {\em IEEE Transactions on Education\/} {\bf 65} 220--242 \urlprefix\url{https://doi.org/10.1109/TE.2022.3144943}

\bibitem{muller2023prperworkforce}
Greinert F, Müller R, Bitzenbauer P, Ubben M~S and Weber K~A 2023 Future quantum workforce: Competences, requirements, and forecasts {\em Physical Review Physics Education Research\/} {\bf 19} 010137 \urlprefix\url{https://link.aps.org/doi/10.1103/PhysRevPhysEducRes.19.010137}

\bibitem{qtmerzeletal}
Merzel A, Bitzenbauer P, Krijtenburg-Lewerissa K, Stadermann K, Andreotti E, Anttila D, Bondani M, Chiofalo M~L~M, Faleti{\v{c}} S, Frans R {\em et~al.\/} 2024 The core of secondary level quantum education: a multi-stakeholder perspective {\em EPJ Quantum Technology\/} {\bf 11} 27 \urlprefix\url{https://doi.org/10.1140/epjqt/s40507-024-00237-x}

\bibitem{kohnle2013}
Kohnle A, Bozhinova I, Browne D, Everitt M, Fomins A, Kok P, Kulaitis G, Prokopas M, Raine D and Swinbank E 2013 A new introductory quantum mechanics curriculum {\em European Journal of Physics\/} {\bf 35} 015001 ISSN 0143-0807 \urlprefix\url{https://doi.org/10.1088/0143-0807/35/1/015001}

\bibitem{rodriguez2020designing}
Rodriguez L~V, van~der Veen J~T, Anjewierden A, van~den Berg E and de~Jong T 2020 Designing inquiry-based learning environments for quantum physics education in secondary schools {\em Physics Education\/} {\bf 55} 065026 \urlprefix\url{https://doi.org/10.1088/1361-6552/abb346}

\bibitem{goorney2024framework}
Goorney S, Bley J, Heusler S and Sherson J 2024 A framework for curriculum transformation in quantum information science and technology education {\em European Journal of Physics\/} {\bf 45} 065702 \urlprefix\url{https://doi.org/10.1088/1361-6404/ad7e60}

\bibitem{bungum2022quantum}
Bungum B and Selst{\o} S 2022 What do quantum computing students need to know about quantum physics? {\em European Journal of Physics\/} {\bf 43} 055706 \urlprefix\url{https://doi.org/10.1088/1361-6404/ac7e8a}

\bibitem{michelini2023research}
Michelini M and Stefanel A 2023 Research studies on learning quantum physics {\em The International Handbook of Physics Education Research: Learning Physics\/} ed Michelini M and Stefanel A (Melville, NY: AIP Publishing LLC) pp 8--1 \urlprefix\url{https://doi.org/10.1063/9780735425477_008}

\bibitem{singh2015review}
Singh C and Marshman E 2015 Review of student difficulties in upper-level quantum mechanics {\em Physical Review Special Topics-Physics Education Research\/} {\bf 11} 020117 \urlprefix\url{https://doi.org/10.1103/PhysRevSTPER.11.020117}

\bibitem{donhauser2024empirical}
Donhauser A, Bitzenbauer P, Qerimi L, Heusler S, K{\"u}chemann S, Kuhn J and Ubben M~S 2024 Empirical insights into the effects of research-based teaching strategies in quantum education {\em Physical Review Physics Education Research\/} {\bf 20} 020601 \urlprefix\url{https://doi.org/10.1103/PhysRevPhysEducRes.20.020601}

\bibitem{beckphoton}
Thorn J~J, Neel M~S, Donato V~W, Bergreen G~S, Davies R~E and Beck M 2004 {Observing the quantum behavior of light in an undergraduate laboratory} {\em American Journal of Physics\/} {\bf 72} 1210--1219 \urlprefix\url{https://doi.org/10.1119/1.1737397}

\bibitem{singh2007comp}
Singh C 2007 Helping students learn quantum mechanics for quantum computing {\em AIP Conf. Proc.\/} {\bf 883} 42--45 \urlprefix\url{https://doi.org/10.1063/1.2508687}

\bibitem{marshman2016ejpphoton}
Marshman E and Singh C 2016 Interactive tutorial to improve student understanding of single photon experiments involving a {M}ach–{Z}ehnder interferometer {\em European Journal of Physics\/} {\bf 37} 024001 ISSN 0143-0807 \urlprefix\url{https://doi.org/10.1088/0143-0807/37/2/024001}

\bibitem{Kohnle_2017}
Kohnle A and Rizzoli A 2017 Interactive simulations for quantum key distribution {\em European Journal of Physics\/} {\bf 38} 035403 \urlprefix\url{https://dx.doi.org/10.1088/1361-6404/aa62c8}

\bibitem{devore2020qkd}
DeVore S and Singh C 2020 Interactive learning tutorial on quantum key distribution {\em Physical Review Physics Education Research\/} {\bf 16} 010126 \urlprefix\url{https://link.aps.org/doi/10.1103/PhysRevPhysEducRes.16.010126}

\bibitem{maries2020mzidouble}
Maries A, Sayer R and Singh C 2020 Can students apply the concept of “which-path” information learned in the context of {M}ach–{Z}ehnder interferometer to the double-slit experiment? {\em American Journal of Physics\/} {\bf 88} 542--550 ISSN 0002-9505 \urlprefix\url{http://dx.doi.org/10.1119/10.0001357}

\bibitem{kiko}
Bista A, Sharma B and Galvez E~J 2021 {A demonstration of quantum key distribution with entangled photons for the undergraduate laboratory} {\em American Journal of Physics\/} {\bf 89} 111--120 \urlprefix\url{https://doi.org/10.1119/10.0002169}

\bibitem{singh2022tpt}
Singh C, Levy A and Levy J 2022 Preparing precollege students for the second quantum revolution with core concepts in quantum information science {\em The Physics Teacher\/} {\bf 60} 639--641 \urlprefix\url{https://doi.org/10.1119/5.0027661}

\bibitem{qtmerzel}
Weissman E~Y, Merzel A, Katz N and Galili I 2024 Keep it secret, keep it safe: teaching quantum key distribution in high school {\em EPJ Quantum Technology\/} {\bf 11} 64 \urlprefix\url{https://doi.org/10.1140/epjqt/s40507-024-00276-4}

\bibitem{hennig2024new}
Hennig F, T{\'o}th K, F{\"o}rster M and Bitzenbauer P 2024 A new teaching-learning sequence to promote secondary school students’ learning of quantum physics using dirac notation {\em Physics Education\/} {\bf 59} 045007 \urlprefix\url{https://doi.org/10.1088/1361-6552/ad353d}

\bibitem{qtbrang}
Brang M, Franke H, Greinert F, Ubben M~S, Hennig F and Bitzenbauer P 2024 Spooky action at a distance? a two-phase study into learners’ views of quantum entanglement {\em EPJ Quantum Technology\/} {\bf 11} 33 \urlprefix\url{https://doi.org/10.1140/epjqt/s40507-024-00244-y}

\bibitem{qthellstern}
Hellstern G, Hettel J and Just B 2024 Introducing quantum information and computation to a broader audience with moocs at openhpi {\em EPJ Quantum Technology\/} {\bf 11} 59 \urlprefix\url{https://doi.org/10.1140/epjqt/s40507-024-00270-w}

\bibitem{qtgoorney}
Goorney S, Sarantinou M and Sherson J 2024 The quantum technology open master: widening access to the quantum industry {\em EPJ Quantum Technology\/} {\bf 11} 7 \urlprefix\url{https://doi.org/10.1140/epjqt/s40507-024-00217-1}

\bibitem{qtmeyercu}
Josephine C~Meyer Gina~Passante S~J~P and Wilcox B~R 2024 Introductory quantum information science coursework at us institutions: content coverage {\em EPJ Quantum Technol.\/} {\bf 11} 16 \urlprefix\url{https://doi.org/10.1140/epjqt/s40507-024-00226-0}

\bibitem{qtsun}
Sun Q, Zhou S, Chen R, Feng G, Cheung K~T, Li J, Hou S~Y and Zeng B 2024 From computing to quantum mechanics: accessible and hands-on quantum computing education for high school students {\em EPJ Quantum Technology\/} {\bf 11} 58 \urlprefix\url{https://doi.org/10.1140/epjqt/s40507-024-00271-9}

\bibitem{hubloch}
Hu P, Li Y, Mong R~S~K and Singh C 2024 Student understanding of the bloch sphere {\em European Journal of Physics\/} {\bf 45} 025705 \urlprefix\url{https://dx.doi.org/10.1088/1361-6404/ad2393}

\bibitem{hucomputing}
Hu P, Li Y and Singh C 2024 Investigating and improving student understanding of the basics of quantum computing {\em Phys. Rev. Phys. Educ. Res.\/} {\bf 20}(2) 020108 \urlprefix\url{https://link.aps.org/doi/10.1103/PhysRevPhysEducRes.20.020108}

\bibitem{lopez2020encrypt}
L{\'o}pez-Incera A, Hartmann A and D{\"u}r W 2020 Encrypt me! a game-based approach to bell inequalities and quantum cryptography {\em European journal of physics\/} {\bf 41} 065702 \urlprefix\url{10.1088/1361-6404/ab9a67}

\bibitem{chhabra2023undergraduate}
Chhabra M and Das R 2023 Undergraduate students’ visualization of quantum mechanical eigenstates and the role of boundary conditions {\em European Journal of Physics\/} {\bf 44} 025702 \urlprefix\url{10.1088/1361-6404/ac9e28}

\bibitem{michelini2022}
Michelini M, Stefanel A and Tóth K 2022 Implementing {Dirac} approach to quantum mechanics in a {Hungarian} secondary school {\em Education Sciences\/} {\bf 12} 606 ISSN 2227-7102 \urlprefix\url{https://doi.org/10.3390/educsci12090606}

\bibitem{schalkers2024explaining}
Schalkers M~A, Dankers K, Wimmer M and Vermaas P 2024 Explaining grover’s algorithm with a colony of ants: a pedagogical model for making quantum technology comprehensible {\em Physics Education\/} {\bf 59} 035003 \urlprefix\url{10.1088/1361-6552/ad2976}

\bibitem{vygotsky}
Vygotsky L~S and Cole M 1978 {\em Mind in Society: The Development of Higher Psychological Processes\/} (Harvard university press) ISBN 0674576292

\bibitem{lave1991situating}
Lave J 1991 Situating learning in communities of practice.  \urlprefix\url{https://doi.org/10.1037/10096-003}

\bibitem{nokes2015better}
Nokes-Malach T~J, Richey J~E and Gadgil S 2015 When is it better to learn together? insights from research on collaborative learning {\em Educational Psychology Review\/} {\bf 27} 645--656 \urlprefix\url{https://doi.org/10.1007/s10648-015-9312-8}

\bibitem{shabani2010vygotsky}
Shabani K, Khatib M and Ebadi S 2010 Vygotsky's zone of proximal development: Instructional implications and teachers' professional development. {\em English language teaching\/} {\bf 3} 237--248 \urlprefix\url{https://doi.org/10.5539/elt.v3n4p237}

\bibitem{ghimire2025reflections}
Ghimire A and Singh C 2025 Reflections of quantum educators on strategies to diversify the second quantum revolution {\em The Physics Teacher\/} {\bf 63} 35--39 \urlprefix\url{https://doi.org/10.1119/5.0223983}

\bibitem{kashyap2025strategies}
Kashyap J~S and Singh C 2025 Strategies educators can use to counter misinformation related to the quantum information revolution {\em Physics Education\/} {\bf 60} 035024 \urlprefix\url{10.1088/1361-6552/adbeb1}

\bibitem{liam}
Doyle L, Seifollahi F and Singh C Do we have a quantum computer? expert perspectives on current state and future prospects

\bibitem{thinkaloud}
Ericsson K~A 2006 Protocol analysis and expert thought: concurrent verbalizations of thinking during experts’ performance on representative tasks {\em The Cambridge Handbook of Expertise and Expert Performance\/} ed Ericsson K~A, Charness N, Feltovich P~J and Hoffman R~R (Cambridge, UK: Cambridge University Press) pp 223--241 \urlprefix\url{https://doi.org/10.1017/CBO9780511816796.013}

\bibitem{patton2002qualitative}
Patton M~Q 2002 {\em Qualitative research and evaluation methods 3rd. ed\/} (Sage publications)

\bibitem{saldana2021coding}
Salda{\~n}a J 2013 {\em The Coding Manual for Qualitative Researchers 2nd ed\/} (London: SAGE publications Ltd)

\bibitem{hedlund2013overview}
Hedlund-de Witt N 2013 An overview and guide to qualitative data analysis for integral researchers {\em Integral Research Center\/} {\bf 1} 76--97

\bibitem{divincenzo}
DiVincenzo D~P 2000 The physical implementation of quantum computation {\em Fortschritte der Physik: Progress of Physics\/} {\bf 48} 771--783 \urlprefix\url{https://doi.org/10.1002/1521-3978(200009)48:9/11<771::AID-PROP771>3.0.CO;2-E}

\bibitem{deutsch}
Deutsch D and Jozsa R 1992 Rapid solution of problems by quantum computation {\em Proceedings of the Royal Society of London. Series A: Mathematical and Physical Sciences\/} {\bf 439} 553--558 \urlprefix\url{https://doi.org/10.1098/rspa.1992.0167}

\bibitem{Shor1994}
Shor P~W 1994 Algorithms for quantum computation: Discrete logarithms and factoring {\em Proceedings of the 35th Annual Symposium on Foundations of Computer Science (FOCS)\/} (IEEE Computer Society) pp 124--134 \urlprefix\url{https://doi.org/10.1109/SFCS.1994.365700}

\bibitem{dancing}
Sutor R~S 2019 {\em Dancing with Qubits: How quantum computing works and how it can change the world\/} (Packt Publishing Ltd)

\bibitem{wong}
Wong T~G 2022 {\em Introduction to classical and quantum computing\/} (Rooted Grove Omaha, Nebraska)

\bibitem{qiskit}
Qiskit \urlprefix\url{https://www.ibm.com/quantum/qiskit}

\bibitem{bennett1984}
Bennett C~H and Brassard G 2014 Quantum cryptography: Public key distribution and coin tossing {\em Theoretical Computer Science\/} {\bf 560} 7--11 ISSN 0304-3975 theoretical Aspects of Quantum Cryptography – celebrating 30 years of BB84 \urlprefix\url{https://www.sciencedirect.com/science/article/pii/S0304397514004241}

\bibitem{quantum_eng_bach_program}
Dzurak A~S, Epps J, Laucht A, Malaney R, Morello A, Nurdin H~I, Pla J~J, Saraiva A and Yang C~H 2022 Development of an undergraduate quantum engineering degree {\em IEEE Transactions on Quantum Engineering\/} {\bf 3} 1--10 \urlprefix\url{https://doi.org/10.1109/TQE.2022.3157338}

\bibitem{sewani2020coherent}
Sewani V~K, Vallabhapurapu H~H, Yang Y, Firgau H~R, Adambukulam C, Johnson B~C, Pla J~J and Laucht A 2020 Coherent control of nv- centers in diamond in a quantum teaching lab {\em American Journal of Physics\/} {\bf 88} 1156--1169 \urlprefix\url{https://doi.org/10.1119/10.0001905}

\bibitem{yang2022observing}
Yang Y, Vallabhapurapu H~H, Sewani V~K, Isarov M, Firgau H~R, Adambukulam C, Johnson B~C, Pla J~J and Laucht A 2022 Observing hyperfine interactions of nv- centers in diamond in an advanced quantum teaching lab {\em American Journal of Physics\/} {\bf 90} 550--560 \urlprefix\url{https://doi.org/10.1119/5.0075519}

\bibitem{simon}
Simon D~R 1997 On the power of quantum computation {\em SIAM Journal on Computing\/} {\bf 26} 1474--1483 \urlprefix\url{https://doi.org/10.1137/S0097539796298637}

\bibitem{vaziranib}
Bernstein E and Vazirani U 1997 Quantum complexity theory {\em SIAM Journal on Computing\/} {\bf 26} 1411--1473 \urlprefix\url{https://doi.org/10.1137/S0097539796300921}

\bibitem{grover}
Grover L~K 2001 {From Schrödinger’s equation to the quantum search algorithm} {\em American Journal of Physics\/} {\bf 69} 769--777 ISSN 0002-9505 \urlprefix\url{https://doi.org/10.1119/1.1359518}

\bibitem{xanadu}
Xanadu \urlprefix\url{https://strawberryfields.ai/}

\bibitem{nielsen2010quantum}
Nielsen M~A and Chuang I~L 2010 {\em Quantum computation and quantum information\/} (Cambridge university press) \urlprefix\url{https://doi.org/10.1017/CBO9780511976667}

\bibitem{mariesnature}
Maries A and Singh C 2024 Towards meaningful diversity, equity and inclusion in physics learning environments {\em Nature Physics\/} {\bf 20} 367--375 \urlprefix\url{https://doi.org/10.1038/s41567-024-02391-6}

\end{thebibliography}

\end{document}